\documentclass[preprint,12pt,sort&compress]{elsarticle}
\usepackage{lineno}
\usepackage{amsmath,bm, amssymb,amsthm,mathrsfs}
\usepackage{mathtools,booktabs}
\usepackage{enumitem}
\usepackage{epstopdf}
\usepackage{float}
\usepackage{epsfig}
\usepackage[caption=false]{subfig}
\usepackage[svgnames]{xcolor}
\usepackage{algorithm}
\usepackage{algpseudocode}
\usepackage{listings}
\usepackage{multicol}
\usepackage{csquotes}
\definecolor{codegreen}{rgb}{0,0.6,0}
\definecolor{codegray}{rgb}{0.5,0.5,0.5}
\definecolor{codepurple}{rgb}{0.58,0,0.82}
\definecolor{backcolour}{rgb}{0.95,0.95,0.92}

\lstdefinestyle{mystyle}{
    backgroundcolor=\color{backcolour},   
    commentstyle=\color{codegreen},
    keywordstyle=\color{magenta},
    numberstyle=\tiny\color{codegray},
    stringstyle=\color{codepurple},
    basicstyle=\ttfamily\scriptsize,
    breakatwhitespace=false,         
    breaklines=true,    captionpos=b,                    
    keepspaces=true,                     
    numbersep=5pt, showspaces=false,                
    showstringspaces=false, showtabs=false, tabsize=2}
\newenvironment{nospaceflalign*}
 {\setlength{\abovedisplayskip}{-0.3cm}\setlength{\belowdisplayskip}{-1pt}%
  \csname flalign*\endcsname}
 {\csname endflalign*\endcsname\ignorespacesafterend}

\newtheorem{thm}{Theorem}

\newdefinition{rmk}{Remark}
\newdefinition{res}{Result}
\newproof{pf}{Proof}
\newproof{pot}{Proof of Theorem \ref{thm2}}

\usepackage{amssymb}

\journal{}

\begin{document}

\begin{frontmatter}

\title{One-Shot Device Testing Data Analysis under Logistic-Exponential Lifetimes with an Application to SEER Gallbladder Cancer Data}
\tnoteref{label1}
 \author{Shanya Baghel$^a$}
 \author{Shuvashree Mondal$^b$\corref{cor1}}
 \ead{shuvasri29@iitism.ac.in}
 \cortext[cor1]{}
 \affiliation{organization={Department of Mathematics and Computing, Indian Institute of Technology (Indian School of Mines) Dhanbad},
             %addressline={},
             %city={},
             postcode={826004},
             state={Jharkhand},
             country={India}}
 %\fntext[label3]{}

\author{}

 \affiliation{organization={Department of Mathematics and Computing, Indian Institute of Technology (Indian School of Mines) Dhanbad},
             %addressline={},
             %city={},
             postcode={826004},
             state={Jharkhand},
             country={India}}

\begin{abstract}
In the literature, the reliability analysis of one-shot devices is found under accelerated life testing in the presence of various stress factors. The application of one-shot devices can be extended to the bio-medical field, where we often evidence that inflicted with a certain disease, survival time would be under different stress factors like environmental stress, co-morbidity, the severity of disease etc.  This work is concerned with a one-shot device data analysis and applies it to SEER Gallbladder cancer data.  The two-parameter logistic exponential distribution is applied as a lifetime distribution.  For robust parameter estimation, weighted minimum density power divergence estimators (WMDPDE) is obtained along with the conventional maximum likelihood estimators (MLE).  The asymptotic behaviour of the WMDPDE and the robust test statistic based on the density power divergence measure are also studied.  The performances of estimators are evaluated through extensive simulation experiments.  Later those developments are applied to SEER Gallbladder cancer data.  Citing the importance of knowing exactly when to inspect the one-shot devices put to the test, a search for optimum inspection times is performed.  This optimization is designed to minimize a defined cost function which strikes a trade-off between the precision of the estimation and experimental cost.  The search is accomplished through the population-based heuristic optimization method Genetic Algorithm.   
\end{abstract}


\begin{highlights}
\item A Robust inference based on weighted density power divergence
\item Elaborated study of the asymptotic property
\item Testing of hypothesis based on a robust statistic
\item Search for optimum inspection times using Genetic algorithm
\item Application to SEER Gallbladder Cancer Data
\end{highlights}


\end{frontmatter}

 %\linenumbers

\section{Introduction}
\label{sec1}
\noindent Applications of one-shot devices prevail in a broad spectrum of life.  One-shot devices stay in a torpid state until activated and are immediately destroyed after the operation.  Amidst the COVID-19 pandemic, extensive use of PPE kits, RT-PCR and Rapid Antigen testing kits etc., is evidenced, which are only for one-time use.  Apart from these, the sugar level testing strips, pregnancy testing kits, fire extinguishers, fuses, fuel injectors, missiles, nuclear weapons, and space probes all come into the category of one-shot devices.  The reliability analysis of such devices becomes quite challenging as the defectiveness of the devices can be discovered only after being tested, and the exact failure times of such devices cannot be recorded accurately in most instances.  Hence it can only be observed whether the device's failure occurs before or after an inspection time, which yields dichotomous data.  With highly reliable devices, typically accelerated life tests (ALTs) are executed to conduct the reliability testing of these devices under a limited time and experimental budget.  With high-stress factors under ALT, reliability is estimated, which is extrapolated to real-life operating situations.  Readers may refer to Wang \cite{WANG2017743}, Xu et al. \cite{xu2015bayesian}, Roy \cite{roy2018bayesian}, Srivastava and Jain \cite{SRIVASTAVA20115786} for the widespread use of ALT in reliability analysis.  The application of one-shot devices can be extended in the bio-medical field, where diseases like cancer and pulmonary infection are often exhibited with different stress factors like environmental stress, lifestyle stress,  co-morbidity and severity of disease etc. \\

\noindent Reviewing the literature for one-shot devices shows that Exponential, Gamma, and Weibull distributions are commonly used as lifetime distributions.  Balakrishnan et al. \cite{balakrishnan2012algorithm} developed an EM algorithm for point estimation under ALT for an exponentially distributed lifetime of one-shot devices. They also compared it with the Bayesian approach using normal prior developed by Fan et al. \cite{fan2009bayesian}.  Balakrishnan et al. \cite{balakrishnan2015bayesian} studied the reliability of one-shot devices in the presence of competing risk factors.  Balakrishnan and Ling \cite{balakrishnan2014gamma}  applied gamma distributions as a lifetime of one-shot devices and provided an inference study.  Balakrishnan and Castilla \cite{balakrishnan2022based} studied the reliability of one-shot devices under log-normal distribution and provided an EM algorithm for estimation purposes under constant stress accelerated life tests.\\

\begin{figure}[htb!]
\begin{center}
\subfloat[$\alpha=1,\beta=0.5$]{\includegraphics[width =0.3\textwidth]{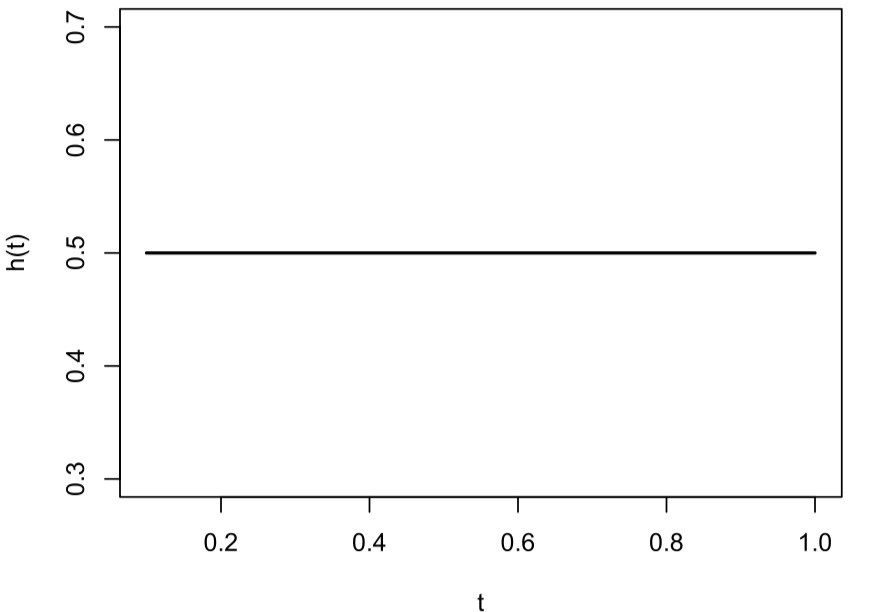}} 
\subfloat[$\alpha=0.8,\beta=0.7$]{\includegraphics[width =0.3\textwidth]{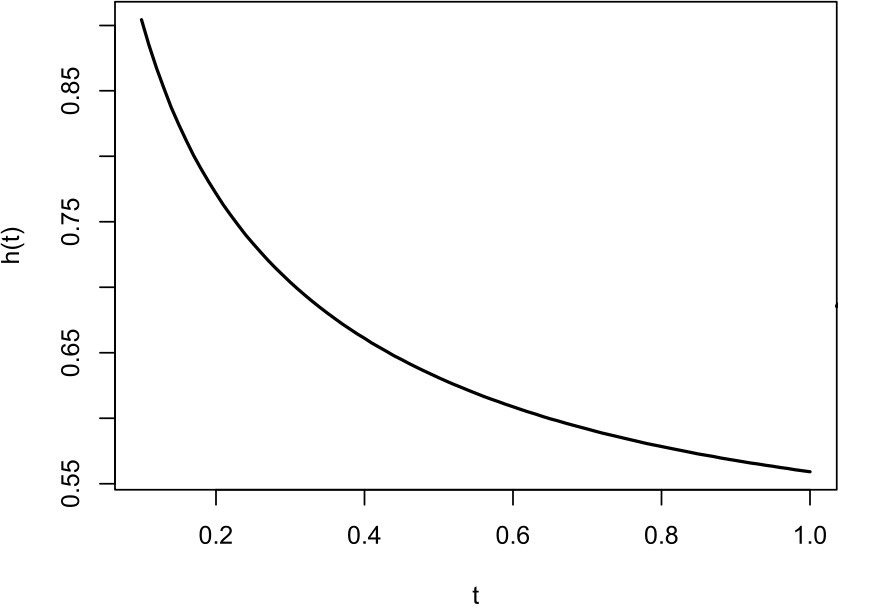}}
\subfloat[$\alpha=4,\beta=0.2$]{\includegraphics[width =0.3\textwidth]{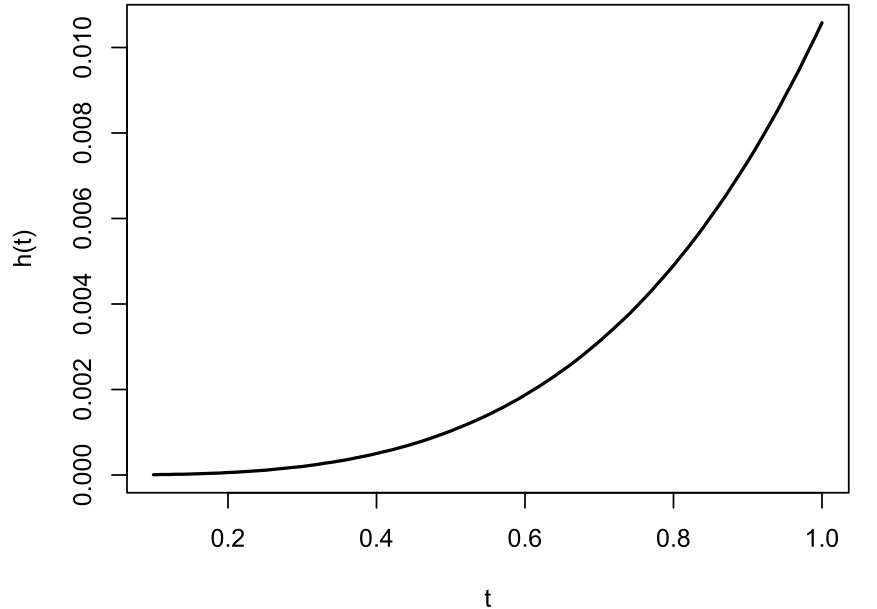}}\\
\subfloat[$\alpha=0.6,\beta=5$]{\includegraphics[width =0.3\textwidth]{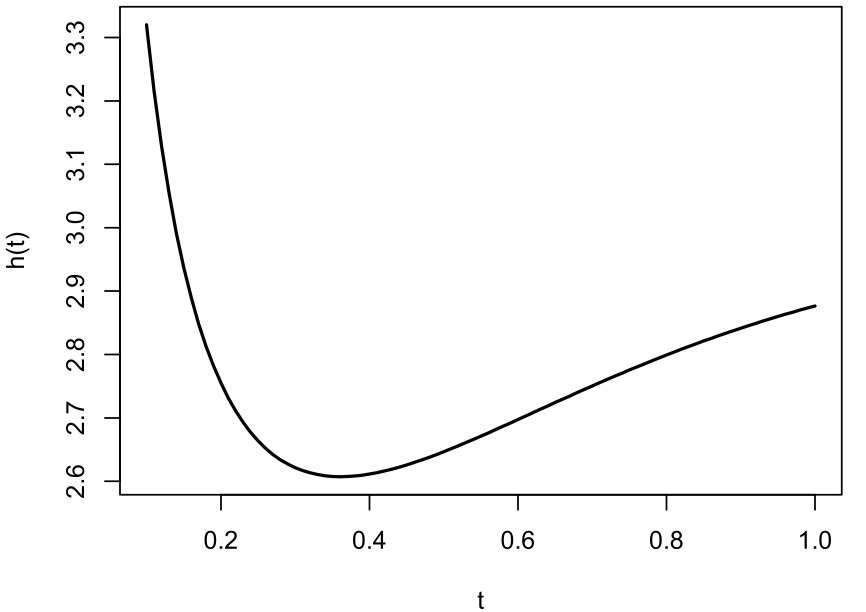}} 
\subfloat[$\alpha=1.5,\beta=4$]{\includegraphics[width =0.3\textwidth]{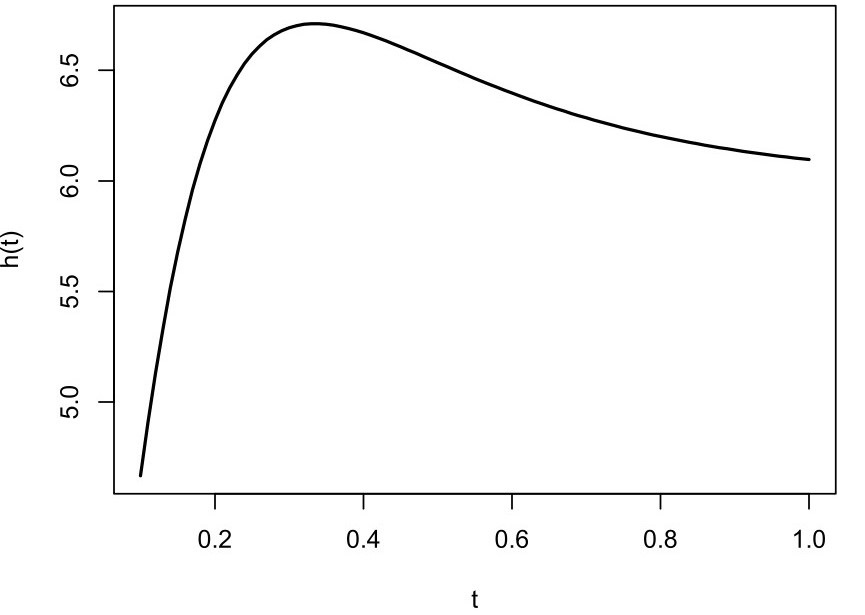}}
\subfloat[$\alpha=2,\beta=0.2$]{\includegraphics[width =0.3\textwidth]{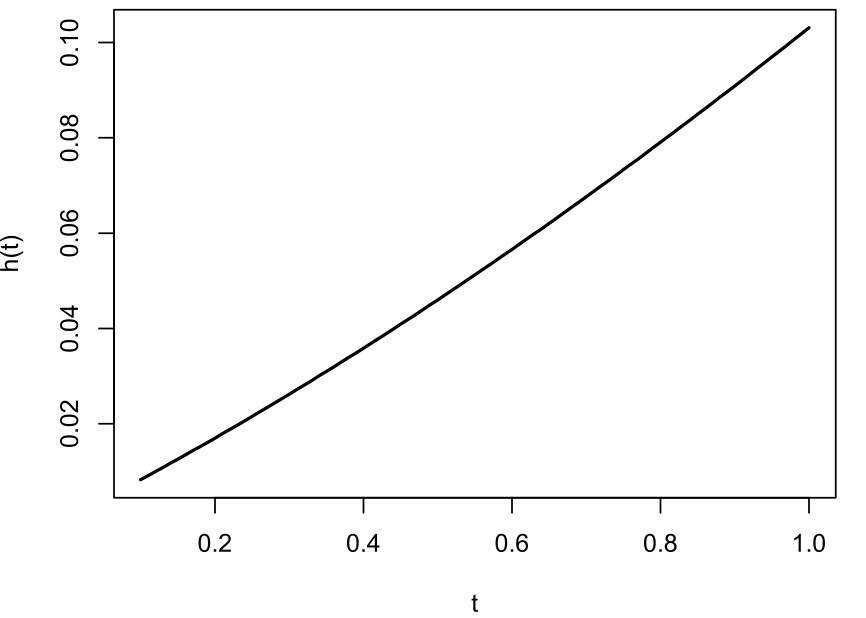}}
\end{center}
\caption{Hazard function of LE distribution for different choices of shape and scale parameters}
\label{fig1}
\end{figure}

\noindent Introduced by Lam and Leemis \cite{lan2008logistic} as a survival distribution, a two-parameter Logistic-Exponential (LE) distribution is chosen in the present work for depicting the lifetime of the one-shot unit model because of its high flexibility regarding the shape of the hazard function. 
 It can be observed that it is the only two parameter distribution which demonstrates five different hazard rate shapes namely, constant, increasing, decreasing, bath-tub and upside-down-bath-tub \cite{ali2020two}.
 Figure \ref{fig1} illustrates the various hazard rate shapes of LE distribution for different values of shape and scale parameters.    As a result, LE distribution can be used as an alternative to some well-known two-parameter models like gamma, log-normal, Weibull, exponentiated exponential, inverse Gaussian, Birnbaum–Saunders distributions \cite{balakrishnan2019birnbaum} and in terms of model fitting, it might work nicely than the stated distributions in some situations.  In the present work, to incorporate the stress factors, the shape and scale parameters of the two-parameter Logistic-Exponential distribution are linked with the stress factors through a log-linear function.\\

\noindent For estimation of the lifetime distribution, though maximum likelihood estimation (MLE) is a prevalent and common method because of its well-known properties such as asymptotic efficiency, consistency, sufficiency, and invariance, it may not possess the property of robustness.  Hence, in this work a robust estimator based on density power divergence measure is used for estimating the lifetime distribution as an alternative.  Minimum density power divergence (MDPD) estimator was first proposed by Basu et.al \cite{basu1998robust}.  In this work, certain weights are imposed on the density power divergence measure and the corresponding estimator is termed as weighted minimum density power divergence estimator (WMDPDE). 
The literature shows that Balakrishnan et al.\cite{balakrishnan2019robust1,balakrishnan2019robust2,balakrishnan2019robust3} observed the robustness of WMDPD estimators for analysing one-shot device testing data under ALT based on gamma, exponential and Weibull lifetime distributions.  
 
\noindent In this work, the asymptotic distribution of the WMDPD estimator is derived.  Inspired by the idea of Basu et al. \cite{basu2013testing}, a testing of hypothesis based on WDPD is also developed in which robust WDPD-based test statistic is presented and an approximation to its power function is determined.  The WMDPDEs and \textcolor{black}{the test statistic} based on the WDPD measure satisfy robustness without compromising efficiency.\\

\noindent Another essential contribution of this work is in the design aspect of the life testing experiment, which consists of finding the optimal inspection times.  Optimal inspection times are the set of inspection times that will optimize certain cost functions, which may be based on maximizing the precision of the estimation, minimizing the experimental cost, or both.  In the literature, it is found that Wu et al. \cite{wu2020optimal} utilized the D-optimality criterion with cost constraints to determine the optimal inspection times.  Ling and Hu \cite{ling2020optimal} minimized the asymptotic variance of MLE under normal operating conditions for Weibull distributions concerning sample allocations and inspection times for determining the optimal designs of simple ALT for one-shot devices.  Balakrishnan and Castilla \cite{balakrishnan2022based} studied optimal design under constant stress accelerated life tests for one-shot devices with budget constraints. \\ 

\noindent Further, a cost function is proposed based on minimizing the determinants of the asymptotic covariance matrix of the WMDPD estimator and the number of expected failures.  Population-based heuristic optimization method Genetic Algorithm (GA) in continuous design space is used in finding optimal inspection times.  The immense popularity of GA lies on the merits of its easy understanding of the concept, ability to converge global or near global optima reasonably well, to avoid the trap of local optima, to work without computation of complex mathematical derivatives,  to deal with various constraints, to use probabilistic transition rules, to show efficiency even when many parameters are involved.\\

\noindent The rest of the work proceeds as follows: Section \eqref{sec2} comprises model description and likelihood function computation.  Density power divergence (DPD) measure with the weighted estimating equations, its asymptotic properties and hypothesis testing, and the power function are discussed in Section \eqref{sec3}.  Section \eqref{sec4} is comprised of an extensive simulation study to assess the performances of the derived estimators.  The cost function is defined in section \eqref{sec5}, and the search for the optimum inspection times applying GA is provided.  In section \eqref{sec6}, outcomes of previous sections are applied to \textcolor{black}{SEER Gallbladder Cancer Data \cite{seer}}.  The study is concluded with a summary discussion in section \eqref{sec7} along with some future insights.

\section{Model Description}\label{sec2}

\noindent In the context of \textcolor{black}{life testing experiment of the one-shot devices,} the devices are set down to $I$ observation groups where the number of devices in $i$th group is $k_i.$ \textcolor{black}{ These devices } are subject to $J$ type of stress factors, quantified by $x_{ij}; j=0,1,2,..., J,$ where $x_{i0}=1.$  In the $i$th observation group, \textcolor{black}{the devices are inspected at a fixed time point say $\tau_i$, and the number of devices fail before $\tau_i$, denoted by $n_i,$ are recorded for $i=1,\ldots, I.$}  Therefore, the observed data can be represented in tabular form as given in Table \eqref{tab1}.

\begin{table}[htb!]
\caption{Model layout and the data representation}
\begin{center}
{\begin{tabular}{lcccccc}
\hline
& & & & \multicolumn{3}{l}{Covariates} \\\cline{5-7}%
Groups & Inspection  & Mice/ & Deaths/ & Stress 1 & \dots & Stress J \\
&Times &Devices &Failures & & \\
\hline
1 & $\tau_1$ & $k_1$&$n_1$ & $x_{11}$ &\dots & $x_{1J}$ \\
	2 & $\tau_2$ & $k_2$&$n_2$ &$x_{21}$ &\dots &$x_{2J}$\\
	\vdots & \vdots & \vdots&\vdots &\vdots &\vdots &\vdots\\
	I&$\tau_I$ &$k_I$ &$n_I$ &$x_{I1}$ &\dots &$x_{IJ}$\\
\hline
\end{tabular}}
\end{center}
\label{tab1}
\end{table}

\noindent In this work, the lifetime $T$ of a device is assumed to follow two parameters Logistic-Exponential distribution with shape parameter $\alpha$ and scale parameter $\lambda$.  The cumulative distribution function and probability density function of $T$ \textcolor{black}{are} given as follows:
\begin{eqnarray*}
	F(t)=&1-(1+(e^{\lambda t}-1)^{\alpha})^{-1}\; ;t>0,\ \alpha,\lambda>0 ,\\
	f(t)=&\frac{\alpha \lambda e^{\lambda t}(e^{\lambda t}-1)^{\alpha -1}}{(1+(e^{\lambda t}-1)^{\alpha})^2};\; t>0,\ \alpha,\lambda >0.
\end{eqnarray*}
Within each observation group, shape and scale parameters are assumed to have log-linear link with the stress factors as follows:
\begin{align*}
	\alpha_i&=\exp\left \{\sum_{j=0}^{J}a_jx_{ij}\right\} \\ 
	\lambda_i&=\exp\left\{\sum_{j=0}^{J}b_jx_{ij}\right\} 
\end{align*}
where $x_{i0}=1$ for $i=1,\ldots,I.$  Denote, $\bm{\theta}=\{a_j,b_j;j=0,...,J\}$ for the model parameters to be estimated.

\noindent The likelihood function on the basis of given observed data can be obtained as
\begin{align}
	L(\bm{\theta})\propto& \prod_{i=1}^{I} F(\tau_i;x_{ij},\bm{\theta})^{n_i}(1-F(\tau_i;x_{ij},\bm{\theta}))^{k_i-n_i}\notag \\
	=&\prod_{i=1}^{I}\left[1-\left\{1+\left(e^{\lambda_i\tau_i}-1\right)^{\alpha_i}\right\}^{-1}\right]^{n_i}\left[\left\{1+\left(e^{\lambda_i\tau_i}-1\right)^{\alpha_i}\right\}^{-1}\right]^{k_i-n_i}.
\end{align}
Therefore, the log-likelihood function without the normalizing constant can be obtained as,

\begin{align}
	ln\;L(\bm{\theta})	\propto&\sum_{i=1}^{I}n_i\;ln\left[\frac{\left(e^{\lambda_i\tau_i}-1\right)^{\alpha_i}}{1+\left(e^{\lambda_i\tau_i}-1\right)^{\alpha_i}}\right]-(k_i-n_i)\;ln\left[1+\left(e^{\lambda_i\tau_i}-1\right)^{\alpha_i}\right]\notag\\
	=&\sum_{i=1}^{I}\left[n_i\alpha_i\;ln\left(e^{\lambda_i\tau_i}-1\right)-k_i\;ln\left\{1+\left(e^{\lambda_i\tau_i}-1\right)^{\alpha_i}\right\}\right]. 
\end{align}
Hence the MLE of $\bm{\theta},$ say $\;\hat{\bm{\theta}}=\{ \hat{a}_j,\hat{b}_j;j=0,...,J\}$ would be derived as 
\begin{equation}
	\hat{\bm{\theta}}=arg\mathop{max}_{\bm{\theta}}ln L(\bm{\theta})
\end{equation}
provided $\sum_{i=1}^{I}n_i>0.$\\

\begin{thm}\label{thm1}
	To obtain the MLE, the set of estimating equations is given as follows,\\
	for the parameter $a_j$,
	\begin{equation}
		\sum_{i=1}^{I}\alpha_i x_{ij}\;ln(e^{\lambda_i\tau_i}-1)\left[n_i-\frac{k_i(e^{\lambda_i\tau_i}-1)^{\alpha_i}}{\left\{1+(e^{\lambda_i\tau_i}-1)^{\alpha_i}\right\}}\right]=0
	\end{equation}
	and for the parameter $b_j$,
	\begin{equation}
		\sum_{i=1}^{I}\frac{\hat{\alpha}_i\lambda_i\tau_ix_{ij}e^{\lambda_i\tau_i}}{(e^{\lambda_i\tau_i}-1)}\left[n_i-\frac{k_i(e^{\lambda_i\tau_i}-1)^{\hat{\alpha}_i}}{\left\{1+(e^{\lambda_i\tau_i}-1)^{\hat{\alpha}_i}\right\}}\right]=0.
	\end{equation}
\end{thm}
\noindent \textcolor{black}{Since MLE is not robust, we discuss a robust method of estimation in the following section.}

\section{Density Power Divergence (DPD) Measure}\label{sec3}

\noindent The minimum divergence estimation method, namely density power divergence for robust parameter estimation, was developed by Basu et al. \cite{basu1998robust}.   \textcolor{black}{Density power divergence} measure between any two probability distributions say, F and G on the same variable with density respectively, f and g can be obtained as:
\begin{align}
	D_{\beta}(g,f )=\int \left\{ f^{\beta+1}(x) -\frac{\beta+1}{\beta}g(x)f^{\beta}(x) 
	+\frac{1}{\beta} g^{\beta+1}(x) \right \} dx 
\end{align}
where \textcolor{black}{$\beta$} $(0\leq\beta\leq 1)$ is \textcolor{black}{called} the tuning parameter.  

\noindent \textcolor{black}{It} is simple to show that, $\quad\mathop{lim}_{\beta \to 0^{+}}D_{\beta}\left(g,f \right)= D_{KL}\left( g\vert f \right).$\\
\textcolor{black}{Here} $D_{KL}(g\vert f )$ is the Kullback-Leibler (KL) divergence measure which is the measure of the information lost when a probability distribution F is used to approximate another probability distribution G on the same variable.\\ 

\noindent Based on the dichotomous data for one-shot devices, the DPD measure between empirical probability distribution,  $\pi_i=\Big(\frac{n_i}{k_i},1-\frac{n_i}{k_i}\Big)$ and theoretical probability distribution,
$p_i=\Big(1-\{1+(e^{\lambda_i \tau_i}-1)^{\alpha_i}\}^{-1}\Big.$, $\Big.\{1+(e^{\lambda_i \tau_i}-1)^{\alpha_i}\}^{-1}\Big)$ for the $ith$ group (for $i=1,\ldots, I$) can be obtained as
\begin{align}
	D_{\beta}( p_i, \pi_i)=& \left[ \frac{1+(e^{\lambda_i\tau_i}-1)^{\alpha_i(\beta+1)}}{\{1+(e^{\lambda_i\tau_i}-1)^{\alpha_i}\}^{\beta+1}}   \right]- \frac{\beta+1}{\beta}\left[\left(\frac{n_i}{k_i}\right)\right.\left(\frac{(e^{\lambda_i\tau_i}-1)^{\alpha_i\beta}}{\{1+(e^{\lambda_i\tau_i}-1)^{\alpha_i}\}^{\beta}}\right) \nonumber\\
	&\left.    +\left(1-\frac{n_i}{k_i}\right)\left(\frac{1}{\{1+(e^{\lambda_i\tau_i}-1)^{\alpha_i}\}^{\beta}}\right) \right] \nonumber \\
	&	+ \frac{1}{\beta}\left[  \left(\frac{n_i}{k_i}\right)^{\beta+1} +\left(1-\frac{n_i}{k_i} \right)^{\beta + 1}\right].
\end{align}

\noindent Considering all I groups and imposing weights  proportional to the group size, the weighted DPD measure with the weights $w_i=\frac{k_i}{K}$ for $i=1,\ldots, I$ where $K=k_1+k_2+...+k_I$ is given as:
\begin{flalign}
	D^w_{\beta}(\bm{\theta})=&	\sum_{i=1}^{I}\frac{k_i}{K}	D_{\beta}(p_i, \pi_i)\notag\\
	=&\sum_{i=1}^{I}\frac{k_i}{K}\left\{\frac{1}{1+\left(e^{\lambda(\tau_i)}-1\right)^{\alpha_i}}\right\}^{\beta}\left[\frac{\left\{1+\left(e^{\lambda_i(\tau_i)}-1\right)^{\alpha_i(\beta+1)}\right\}}{\left\{1+\left(e^{\lambda_i(\tau_i)}-1\right)^{\alpha_i}\right\}}\right.\notag\\
	&\qquad\qquad\left.-\frac{\beta+1}{\beta}\left\{\frac{n_i}{k_i}\left(e^{\lambda_i(\tau_i)}-1\right)^{\alpha_i\beta}+\left(1-\frac{n_i}{k_i}\right)\right\}\right]\notag\\
	&\qquad\qquad+\frac{1}{\beta}\left\{\left(\frac{n_i}{k_i}\right)^{\beta+1}+\left(1-\frac{n_i}{k_i}\right)^{\beta+1}\right\}\label{8}
 &&
\end{flalign}

\noindent The relationship between the likelihood function and the weighted DPD measure can be established as follows.
\begin{align*}
	\mathop{lim}_{\beta \to 0^{+}}D^w_{\beta}(\bm{\theta})
	=&C-\frac{1}{K}\;ln L(\bm{\theta})
\end{align*}
where $C$ being a constant which is $\bm{\theta}$ independent.\\
\noindent Weighted Minimum density power divergence estimators (WMDPDE) for $\bm{\theta}$ can be defined as follows.
\begin{align}
	\hat{\bm{\theta}}_{\beta}=&arg\mathop{min}_{\bm{\theta}}D^w_{\beta}(\bm{\theta})\label{9}
\end{align}
\textcolor{black}{In parameter estimation, the tuning parameter $\beta$ plays an important role.  It brings a balance between the efficiency and robustness of the WMDPD estimator.}

\begin{thm}\label{thm2}
	\noindent The set of estimating equations to minimize WMDPD measure is given as follows.
	\begin{flalign}
		\sum_{i=1}^{I}k_i&\left[\left\{\left(1-(1+(e^{\lambda_i \tau_i}-1)^{\alpha_i})^{-1}\right)^{\beta-1}+(1+(e^{\lambda_i \tau_i}-1)^{\alpha_i})^{-(\beta-1)}\right\}\right.\notag\\
	&	\left.\left\{\left(1-(1+(e^{\lambda_i \tau_i}-1)^{\alpha_i})^{-1}\right)-\frac{n_i}{k_i}\right\}\right]
		\left[\frac{\partial F(\tau_i;x_{ij},\bm{\theta})}{\partial\bm{\theta}}\right]=0_{2J}.
		&&
	\end{flalign}
	with $\bm{\theta}=(a_1,..,a_j,..,a_J,b_1,..,b_j,..,b_J)^{'}$ and 
	\begin{flalign}
	\frac{\partial F(\tau_i;x_{ij},\bm{\theta})}{\partial a_j}&=\frac{\alpha_ix_{ij}(e^{\lambda_i\tau_i}-1)^{\alpha_i}\,ln(e^{\lambda_i\tau_i}-1)}{\left\{1+(e^{\lambda_i\tau_i}-1)^{\alpha_i}\right\}^2}\notag\\
	\frac{\partial F(\tau_i;x_{ij},\bm{\theta})}{\partial b_j}&=\frac{\alpha_i\lambda_i\tau_i\,x_{ij}\,e^{\lambda_i\tau_i}(e^{\lambda_i\tau_i}-1)^{\alpha_i-1}}{\left\{1+(e^{\lambda_i\tau_i}-1)^{\alpha_i}\right\}^2}\notag
	&&
	\end{flalign}
For $\beta$ \textcolor{black}{tends to} 0, WMDPDE \textcolor{black}{converges} to MLE.
\end{thm}

\subsection{\textbf{Asymptotic Property}}
\noindent Here, we present the asymptotic distribution of the weighted minimum density power divergence estimator under the logistic exponential distribution based on the failure count data.
\begin{thm}\label{thm3}
	Suppose $\bm{\theta}_0$ is the true value of the parameter $\bm{\theta},$  when $k_i \rightarrow \infty$ and $\lim_{k_i \to \infty}\frac{k_i}{K}$ finite for all $i=1,\ldots, I$
	\begin{align}
		\sqrt{K}(\hat{\bm{\theta}}_{\beta}-\bm{\theta}_0)\xrightarrow[K\to \infty]{\mathscr{L}} N\left(0_{2J},J_{\beta}^{-1}(\bm{\theta}_0)K_{\beta}(\bm{\theta}_0)J_{\beta}^{-1}(\bm{\theta}_0)\right)
	\end{align}
\end{thm}
\begin{pf}
Given in Appendix B.
\end{pf}

\subsection{\textbf{Testing of hypothesis based on weighted Density Power Divergence (WDPD) }}

\noindent Testing of statistical hypothesis is of fundamental importance \textcolor{black}{in} inferential analysis.  Due to the lack of robustness of MLE, tests based on MLE may not be able to perform well in the presence of \textcolor{black}{outliers in the data set.}  Here \textcolor{black}{we present} a Robust WDPD-based test statistic for testing a simple \textcolor{black}{null} hypothesis based on the idea of Basu et al. \cite{basu2013testing}.  The null and alternative hypoth\textcolor{black}{eses} \textcolor{black}{are} given as follows.
$$H_0:\bm{\theta}=\bm{\theta}_0\qquad\text{against}\qquad H_1:\bm{\theta}\ne\bm{\theta}_0$$
where $\bm{\theta}=\{a_j,b_j;j=1,2,\dots,J\}$.\\
The robust WDPD-based test statistic is defined as,
\begin{equation}
  \Lambda_{\beta}(\hat{\bm{\theta}}_{\beta},\bm{\theta}_0) =2KD^w_{\beta}(\hat{\bm{\theta}}_{\beta},\bm{\theta}_0) 
\end{equation}
where,
\begin{flalign*}
D^w_{\beta}(\hat{\bm{\theta}}_{\beta},\bm{\theta}_0) &=\sum_{i=1}^{I}\frac{k_i}{K}\left[\{F^{\beta+1}_i(\bm{\theta}_0)+\bar{F}^{\beta+1}_i(\bm{\theta}_0)\}-\frac{\beta+1}{\beta}\left\{F_i(\hat{\bm{\theta}}_{\beta})F^{\beta}_i(\bm{\theta}_0)\right.\right.\\
&\qquad\left.\left.+\bar{F}_i(\hat{\bm{\theta}}_{\beta})\bar{F}^{\beta}(\bm{\theta}_0)\right\}+\frac{1}{\beta}\left\{F_i^{\beta+1}(\hat{\bm{\theta}}_{\beta})+\bar{F}_i^{\beta+1}(\hat{\bm{\theta}}_{\beta})\right\}\right].
&&
\end{flalign*}
Applying Taylor series expansion of second order around $\bm{\theta}=\bm{\theta}_0$ at $\bm{\theta}=\hat{\bm{\theta}}_{\beta}, $ \textcolor{black}{$\Lambda_{\beta}(\hat{\bm{\theta}}_{\beta},\bm{\theta}_0)$  can be expressed as,}
\begin{flalign}
\Lambda_{\beta}(\hat{\bm{\theta}}_{\beta},\bm{\theta}_0)&=2K\left[D^w_{\beta}(\bm{\theta}_{0},\bm{\theta}_0)+(\hat{\bm{\theta}}_{\beta}-\bm{\theta}_0)^{T}\nabla D^w_{\beta}(\bm{\theta}_0,\bm{\theta}_0)\right. \notag\\
&\qquad\qquad+\left.
\frac{1}{2}(\hat{\bm{\theta}}_{\beta}-\bm{\theta}_0)^{T} \nabla^2 D^w_{\beta}(\bm{\theta}_{0},\bm{\theta}_0)(\hat{\bm{\theta}}_{\beta}-\bm{\theta}_0)\right].\notag\\
\text{When $H_0$ is true, } \notag\\ \Lambda_{\beta}(\hat{\bm{\theta}}_{\beta},\bm{\theta}_0)&=\sqrt{K}(\hat{\bm{\theta}}_{\beta}-\bm{\theta}_0)^{T}\nabla^2 D^w_{\beta}(\bm{\theta}_{0},\bm{\theta}_0)\sqrt{K}(\hat{\bm{\theta}}_{\beta}-\bm{\theta}_0)\\
&\qquad(\text{since}\;\;D^w_{\beta}(\bm{\theta}_{0},\bm{\theta}_0)=0\,,\;\nabla D^w_{\beta}(\bm{\theta}_{0},\bm{\theta}_0)=0)\notag
&&
\end{flalign}
\textcolor{black}{ For further development, the following result is presented.}
\begin{res}
Let $\bm{X}\sim N_p(0,\Sigma)$ and $\bm{A}$ be a p$\times$p real symmetric matrix then $\bm{X^{'}AX}$ can be expressed as 
\begin{flalign*}
\bm{X^{'}AX} &=\bm{X^{'}\Sigma^{-1/2}\Sigma^{1/2}A\Sigma^{1/2}\Sigma^{-1/2}X}\\
&=\bm{X^{'}\Sigma^{-1/2}P\Lambda P^{'}\Sigma^{-1/2}X}\\
&=\bm{W^{'}\Lambda W}.
&&
\end{flalign*}
Here, $\bm{P}$ is an orthogonal matrix containing the eigen vectors of $\bm{\Sigma^{1/2}A\Sigma^{1/2}}.$  $\bm{\Lambda}$ is a diagonal matrix with diagonal elements $\lambda_l$ which are eigen values of $\bm{\Sigma^{1/2}A\Sigma^{1/2}}$ and $\bm{W}=\bm{P^{'}\Sigma^{-1/2}X}$. 
\end{res}
As $\bm{W\sim N(0, I)},$ therefore, the distribution of $\bm{X^{'}AX}$ is same as the distribution of $\sum_{l=1}^{r}\lambda_lW^2_l$, where  $W_1, \ldots, W_r$ are the independent standard normal variables. Here, $r=rank\bm{(\Sigma^{1/2}A\Sigma^{1/2})}=rank\bm{(A\Sigma)}$ and $\lambda_is$ are the non-zero eigenvalues of $\bm{A\Sigma}$.\\

\noindent Using the above result, the asymptotic distribution of test statistic $\Lambda_{\beta}(\hat{\bm{\theta}}_{\beta},\bm{\theta}_0)$ can be described by $\sum_{l=1}^{r}\lambda^{(\beta)}_lW^2_l$, where $\lambda^{(\beta)}_ls$ are the non-zero eigen values of $\bm{(\nabla^2D^w_{\beta}(\bm{\theta}_0, \bm{\theta}_0) \Sigma_{\beta}(\bm{\theta}_0)})$ and $\bm{\Sigma_{\beta}(\bm{\theta}_0)}$= $\bm{J^{-1}_{\beta}(\bm{\theta}_0)K_{\beta}(\bm{\theta}_0)}$ $\bm{J^{-1}_{\beta}(\bm{\theta}_0)}$ with $r=rank\bm{(\nabla^2D^w_{\beta}}$ $\bm{(\bm{\theta}_0, \bm{\theta}_0 )\Sigma_{\beta}(\bm{\theta}_0) )}$.\\
\noindent Further, let us define,
$$\Lambda^{*}_{\beta}(\hat{\bm{\theta}}_{\beta},\bm{\theta}_0)=\frac{\Lambda_{\beta}(\hat{\bm{\theta}}_{\beta},\bm{\theta}_0)}{\lambda^{(\beta)}_{max}}\leq \sum_{l=1}^{r}W^2_l$$ where $\lambda^{(\beta)}_{max}=max(\lambda^{(\beta)}_l;l=1,2,\dots,r)$.  As, $\sum_{l=1}^{r}W^2_l\sim \chi^2_{(r)}$, $H_0$ is rejected when $\Lambda^{*}_{\beta}(\hat{\bm{\theta}}_{\beta},\bm{\theta}_0)\geq \chi^2_{(r,1-\alpha)}$ where $\chi^2_{(r,1-\alpha)}$ is the upper $(1-\alpha)$ quantile point of $\chi^2_{(r)}$.

\subsection{\textbf{Power Function of WDPD Based Test}}
\noindent In this section, we \textcolor{black}{study} the power function of WDPD based test. The alternative hypothesis is set as $H_{1,K}:\bm{\theta}_{K}=\bm{\theta}_0+K^{-1/2}\mathbf{d}$, where $\mathbf{\bm{\theta}}_K\in\mathbf{\bm{\theta}}\subset\bm{\mathbb{R}^{2J}}$ and $\mathbf{d}$ is a fixed vector in $\bm{\mathbb{R}^{2J}}$.  \textcolor{black}{Further} $\sqrt{K}(\bm{\hat{\theta}}_{\beta}-\bm{\theta}_0)$ can be written as,
\begin{flalign*}
\sqrt{K}(\bm{\hat{\theta}}_{\beta}-\bm{\theta}_0)&=\sqrt{K}(\bm{\hat{\theta}_{\beta}}-\bm{\theta}_K)+\bm{d}\\
\qquad\text{Under }H_{1,K},&\\
\sqrt{K}(\bm{\hat{\theta}}_{\beta}-\bm{\theta}_K)&\xrightarrow[K\to \infty]{\mathscr{L}} N\left(0_{\bm{2J}},\bm{\Sigma}_{\beta}(\bm{\theta}_K)\right)\quad\text{and therefore}\\
\sqrt{K}(\bm{\hat{\theta}}_{\beta}-\bm{\theta}_0)&\xrightarrow[K\to \infty]{\mathscr{L}} N\left(\bm{d},\bm{\Sigma}_{\beta}(\bm{\theta}_K)\right)
&&
\end{flalign*}
The establishment of the asymptotic distribution of $\Lambda_{\beta}(\hat{\bm{\theta}}_{\beta},\bm{\theta}_0)$ under $H_{1,K}$ is based on the Corollary 2.2 of Dik and De Gunst \cite{dik1985distribution} \textcolor{black}{which is} given as follows.\\

\begin{res}
Let $\bm{X\sim N_p(\mu,\Sigma)}$, $\bm{A}$ be a real-symmetric non-negative definite matrix of order p, $r=rank(\bm{\Sigma A\Sigma}), r\geq 1$ and $\lambda_1,\dots,\lambda_r$ be the positive eigenvalues of $\bm{A\Sigma}$. Then the distribution of $\bm{X^{'}AX}$ is equivalent to the distribution of $\sum_{l=1}^{r}\lambda_l(W_l+v_l)^2+\bm{\Psi}$, where $W_1,\dots,W_r$ are independent standard normal variables and $\bm{v}=\bm{\Lambda^{-1}P^{'}S^{'}A\mu}$, $\bm{\Psi=\mu^{'}A\mu-v^{'}\Lambda v}$.  Here, $\bm{S}$ be any square root of $\bm{\Sigma}$ and $\bm{S^{'}AS=P\Lambda P^{'}}$ where $\bm{\Lambda}$ is a diagonal matrix with diagonal elements $\lambda_1,\dots,\lambda_r$ which are positive eigenvalues of $\bm{S^{'}AS}$ and $\bm{P}$ is an orthogonal matrix with column vectors being the eigenvectors of the corresponding eigenvalues.
\end{res}

\noindent Using the above result, the asymptotic distribution of $\Lambda_{\beta}(\hat{\bm{\theta}}_{\beta},\bm{\theta}_0)$ under $H_{1,K}$ is equivalent with the distribution of $\sum_{l=1}^{r}\lambda^{(\beta)}_l(\bm{\theta}_0)(W_l+v_l)^2+\bm{\Psi}$ where, \textcolor{black}{$\lambda^{(\beta)}_1(\bm{\theta_0}),\dots,\lambda^{(\beta)}_r(\bm{\theta_0}) $ are the positive eigenvalues of $\bm{({\nabla^2D^w_{\beta}(\theta_0,\theta_0)\Sigma_{\beta}(\theta_0)})}$,}
\color{black}
\begin{nospaceflalign*}
&W_l \mathop\sim N(0,1) \ \text{independently for} \ l=1, \ldots,r,\\
&\bm{v}=\bm{\Lambda^{-1}P^{'}S^{'}\nabla^2D^w_{\beta}(\theta_0,\theta_0)d} \ \text{where} \ \bm{v}=(v_1,\dots,v_r)^{'},\\
&\bm{\Psi}=\bm{d^{'}\nabla^2D^w_{\beta}(\theta_0,\theta_0)d-v^{'}\Lambda v},\\
& \bm{S}\; \text{is square root of}\;\bm{\Sigma_{\beta}(\theta_0)}, \\
 & \bm{\Lambda}=diag(\lambda^{(\beta)}_1(\bm{\theta}_0),\dots,\lambda^{(\beta)}_r(\bm{\theta}_0)),\\
&\lambda^{(\beta)}_l(\bm{\theta}_0) \ \text{for} \  l=1,\ldots,r, \ \text{are the positive eigenvalues of} \ \bm{(S^{'}\nabla^2D^w_{\beta}(\theta_0,\theta_0)S)}\\
& \text{and} \  \bm{P} \ \text{is an orthogonal matrix with column vectors being the eigenvectors of}\\
& \text{the corresponding eigenvalues.}
&&
\end{nospaceflalign*}

\color{black}

\noindent Though the asymptotic distribution of $\Lambda_{\beta}(\hat{\bm{\theta}}_{\beta},\bm{\theta}_0)$ under $H_{1, K}$ using the above method is very informative, yet it does not help determine the power function due to its complex nature. The alternative approach for determining the power function approximation is given as follows.\\
\noindent The first order Taylor expansion of $D^w_{\beta}(\hat{\bm{\theta}}_{\beta},\bm{\theta}_0)$ under $\bm{\theta^{*}}$, $\bm{\theta^{*}}\neq \bm{\theta}_0$ is given as,
$$
D^w_{\beta}(\hat{\bm{\theta}}_{\beta},\bm{\theta}_0)=D^w_{\beta}(\bm{\theta^{*}},\bm{\theta}_0)+\bm{A^{'}_{\beta}}(\bm{\hat{\theta}}_{\beta}-\bm{\theta}^{*}),\,\text{where }\,\bm{A_{\beta}}=\nabla D^w_{\beta}(\bm{\theta}^{*},\bm{\theta}_0) 
$$
As
$$
\sqrt{K}(\bm{\hat{\theta}}_{\beta}-\bm{\theta}^{*})\xrightarrow[K\to \infty]{\mathscr{L}} N\left(0_{\bm{2J}},\bm{\Sigma}_{\beta}(\bm{\theta}^{*})\right)
$$
Then,$\;\sqrt{K}(D^w_{\beta}(\hat{\bm{\theta}}_{\beta},\bm{\theta}_0)-D^w_{\beta}(\bm{\theta}^{*},\bm{\theta}_0))$ and $\bm{A^{'}_{\beta}}\sqrt{K}(\bm{\hat{\theta}}_{\beta}-\bm{\theta}^{*})$ have the same asymptotic distribution and 
$$
\sqrt{K}(D^w_{\beta}(\hat{\bm{\theta}}_{\beta},\bm{\theta}_0)-D^w_{\beta}({\bm{\theta}}^{*},\bm{\theta}_0))\xrightarrow[K\to \infty]{\mathscr{L}} N\left(0_{\bm{2J}},\bm{\Sigma}^{*}_{\beta}(\bm{\theta}^{*})\right)
$$
where, $\bm{\Sigma}^{*}_{\beta}(\bm{\theta}^{*})=\bm{A}^{'}_{\beta}\bm{\Sigma}_{\beta}(\bm{\theta}^{*})\bm{A}_{\beta}$.\\

\noindent Therefore the power function can be obtained as 
\begin{flalign}
\bm{\pi}^{(\beta)}_{K,\alpha}(\bm{\theta}^{*})&=Pr[2KD^w_{\beta}(\hat{\bm{\theta}}_{\beta},\bm{\theta}_0)>c^{(\beta)}_{\alpha}]\notag\\
&=Pr\left[\frac{\sqrt{K}(D^w_{\beta}(\hat{\bm{\theta}}_{\beta},\bm{\theta}_0)-D^w_{\beta}(\bm{\theta}^{*},\bm{\theta}_0))}{\bm{\Sigma}^{*}_{\beta}(\bm{\theta}^{*})}\right.\notag\\
&\qquad\qquad \left.>\frac{\sqrt{K}}{\bm{\Sigma}^{*}_{\beta}(\bm{\theta}^{*})}\left(\frac{c^{(\beta)}_{\alpha}}{2K}-D^w_{\beta}(\bm{\theta}^{*},\bm{\theta}_0)\right)\right]\notag\\
\bm{\pi}^{(\beta)}_{K,\alpha}(\theta^{*})&=1-\Phi\left[\frac{\sqrt{K}}{\bm{\Sigma}^{*}_{\beta}(\bm{\theta}^{*})}\left(\frac{c^{(\beta)}_{\alpha}}{2K}-D^w_{\beta}(\bm{\theta}^{*},\bm{\theta}_0)\right)\right]
&&
\end{flalign}
where, $\Phi(x)$ is the standard normal distribution function and $c^{(\beta)}_{\alpha}$ is $(1-\alpha)$ percentile of the distribution of $\Lambda_{\beta}(\hat{\bm{\theta}},\bm{\theta}_0)$ under $H_0$.

\section{Simulation Experiment}\label{sec4}
\noindent In this section, a simulated environment is created using the Markov Chain Monte Carlo simulations based on 1000 generations and performances of MLE and WMDPDEs are observed.

\noindent The lifetimes of the one-shot devices are considered to follow the Logistic-Exponential distribution.  An accelerated life test is conducted under three testing groups, with three inspection times.  A different number of devices are put to the test in each group which are subjected to two stress factors.  The layout summaries are given in Table \eqref{tab2}.

\begin{table}[htb!]
\caption{Layout of one-shot device ALT design for simulation}
\begin{center}
{\begin{tabular}{lccccc}
\hline%
& & & & \multicolumn{2}{l}{Covariates} \\\cline{5-6}%
Groups & Inspection Times & Devices & Failures & Stress 1 & Stress 2 \\
\hline
	1 & $1.00$ & $\color{black}{15}$&$n_1$ & $0.2$ & $0.4$ \\
			2 & $1.00$ & $\color{black}{20}$&$n_2$ &$0.3$ & $0.6$\\
			3&$1.00$&$\color{black}{25}$ &$n_3$ &$0.4$ & $0.8$\\
\hline
\end{tabular}}
\end{center}
\label{tab2}
\end{table}

\noindent The three sets of model parameters are taken so that the results can be analysed with at least three different viewpoints and the performances of the estimators can be assessed.  The generated data has been contaminated in three ways, and the results are analysed.  Different sets of model parameters for pure ALT data \textcolor{black}{and contamination scheme} are given in Table \eqref{tab3}.
\begin{table}[htb!]
\color{black}
\caption{Model parameters to generate data}
\begin{center}
{\begin{tabular}{lcccc}
\hline
\textbf{S.No.}& $\;\mathbf{a_1}\;$ & $\;\mathbf{a_2}\;$ & $\;\mathbf{b_1}\;$ & $\;\mathbf{b_2}\;$\\
\hline
\textbf{Pure Data} & & & & \\
$\bm{\theta}_1$ & $0.2$ & $-0.6$&$-0.2$ &$0.4$ \\
$\bm{\theta}_2$ & $0.4$ & $0.3$&$-0.1$& $-0.2$\\
$\bm{\theta}_3$&$-0.06$&$-0.06$ &$0.4$ & $-0.1$\\
\hline
\textbf{Contamination} & & & & \\
$\bm{\theta}_1$ &$a_1+0.02$ &$a_2-0.09$ &$b_1-0.07$ &$b_2-0.01$\\
$\bm{\theta}_2$ &$a_1-0.07$ &$a_2+0.06$ &$b_1-0.05$ &$b_2+0.03$\\
$\bm{\theta}_3$&$a_1+0.00 $ &$a_2-0.07$ &$b_1+0.07$ &$b_2+0.05$\\
\hline
\end{tabular}}
\end{center}
\label{tab3}
\end{table}
\textcolor{black}{For the given data, the Coordinate-Descent method is used to obtain MLEs and WMDPDEs.  The steps of the Coordinate-Descent algorithm are given in the Algorithm \eqref{alg1}}.\\
\begin{algorithm}[htb!]
\color{black}
\caption{\textcolor{black}{Coordinate-Descent Algorithm}}\label{alg1}
\begin{itemize}[noitemsep]
\item Choose initial value of $\bm{\theta}=(a_1,a_2, b_1, b_2)$ say $\bm{\theta}_0=(a^{(0)}_1,a^{(0)}_2,b^{(0)}_1,b^{(0)}_2)$.
\item Let at $t$th iteration for $t=0, 1,2,\ldots,$  the estimate of $\bm{\theta}$ be $\bm{\theta}^t=(a^{(t)}_1,a^{(t)}_2,b^{(t)}_1,b^{(t)}_2)$.  Then at $t+1$th iteration, the estimate can be derived as,\\
		$
		a^{(t+1)}_1=a^{(t)}_1 - h\frac{\partial  H(a^{(t)}_1,a^{(t)}_2,b^{(t)}_1,b^{(t)}_2) }{\partial a_1}\\
		a^{(t+1)}_2=a^{(t)}_2 - h\frac{\partial  H(a^{(t+1)}_1,a^{(t)}_2,b^{(t)}_1,b^{(t)}_2) }{\partial a_2}\\
		b^{(t+1)}_1=  b^{(t)}_1 - h\frac{\partial  H(a^{(t+1)}_1,a^{(t+1)}_2,b^{(t)}_1,b^{(t)}_2) }{\partial b_1}\\
		b^{(t+1)}_2 = b^{(t)}_2- h\frac{\partial  H(a^{(t+1)}_1,a^{(t+1)}_2,b^{(t+1)}_1,b^{(t)}_2) }{\partial b_2}\\
		$
		where $H=- \ln L(\bm{\theta})$ in case of MLE and  $H=D^{w}_{\beta}(\bm{\theta})$ in case of WMDPDE where $h$ is the learning rate.  The value of the learning rate is chosen as $h=0.01.$
\item The process continues until  
		$max( \vert a^{(t+1)}_1 -a^{(t)}_1\vert,  \vert a^{(t+1)}_2 -a^{(t)}_2\vert,  \vert b^{(t+1)}_1 -b^{(t)}_1\vert,  \vert b^{(t+1)}_2 -b^{(t)}_2\vert )$ is less than some pre-specified threshold value and if it is satisfied the final estimate is obtained as $\bm{\theta}^{t+1}=(a^{(t+1)}_1, a^{(t+1)}_2, b^{(t+1)}_1, b^{(t+1)}_2 ).$
\end{itemize}
\end{algorithm}

\begin{figure}[htb!]
\color{black}
\begin{center}
\subfloat[Group 1]{\includegraphics[width =0.5\textwidth]{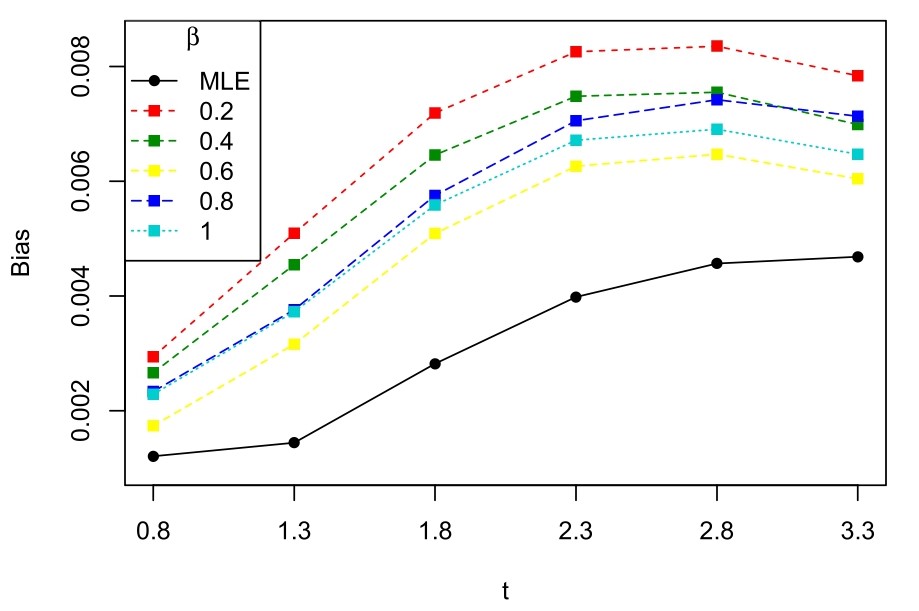}} 
\subfloat[Group 1]{\includegraphics[width =0.5\textwidth]{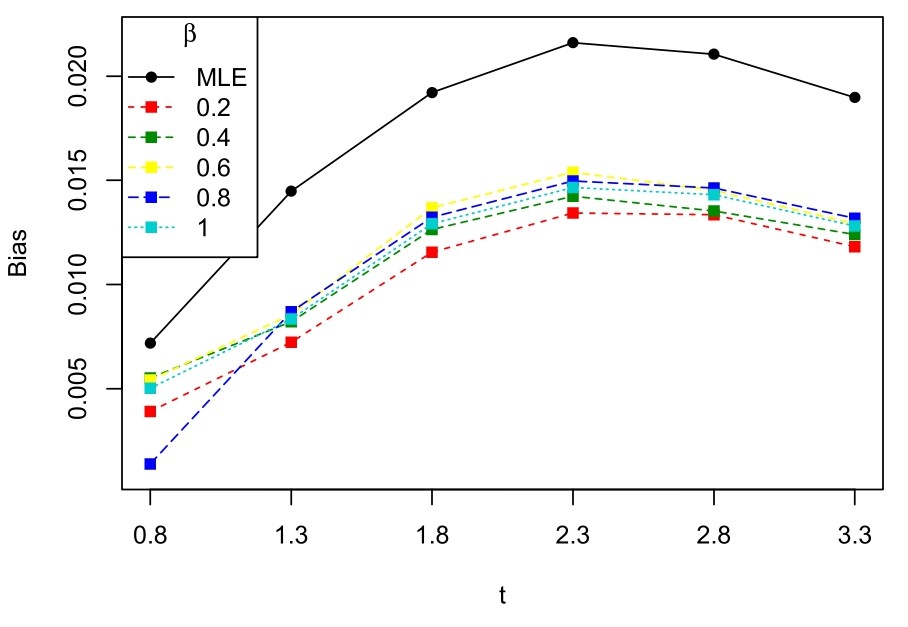}}\\
\subfloat[Group 2]{\includegraphics[width =0.5\textwidth]{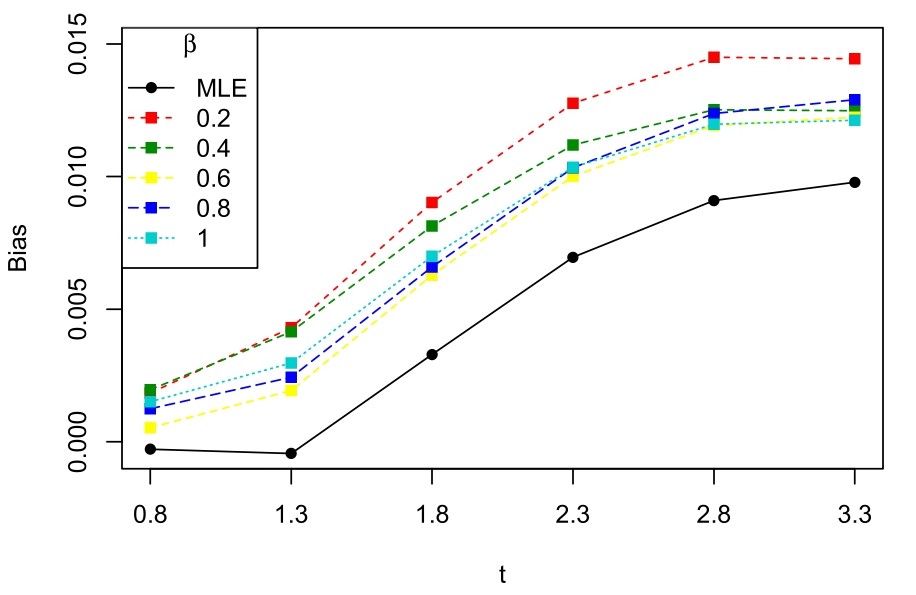}}
\subfloat[Group 2]{\includegraphics[width =0.5\textwidth]{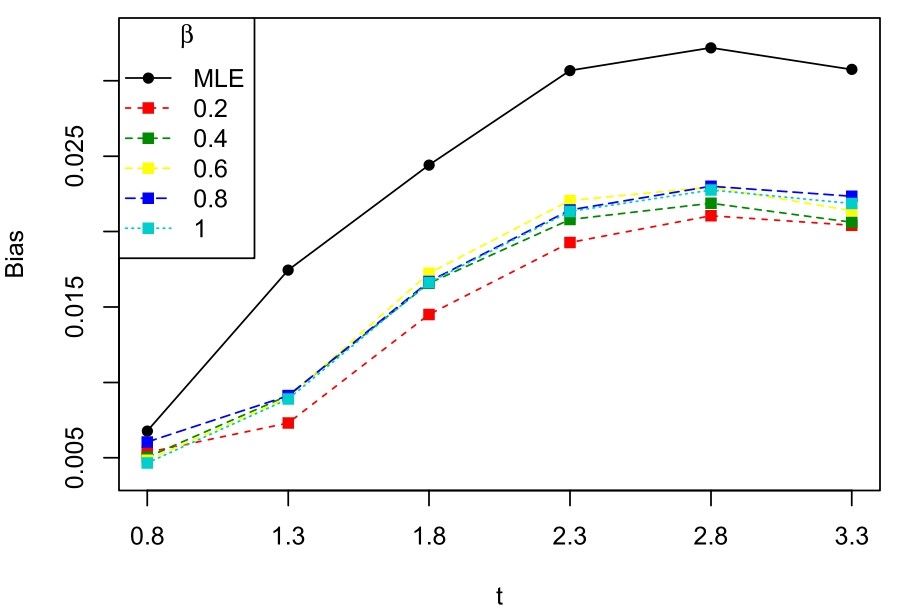}} \\
\subfloat[Group 3]{\includegraphics[width =0.5\textwidth]{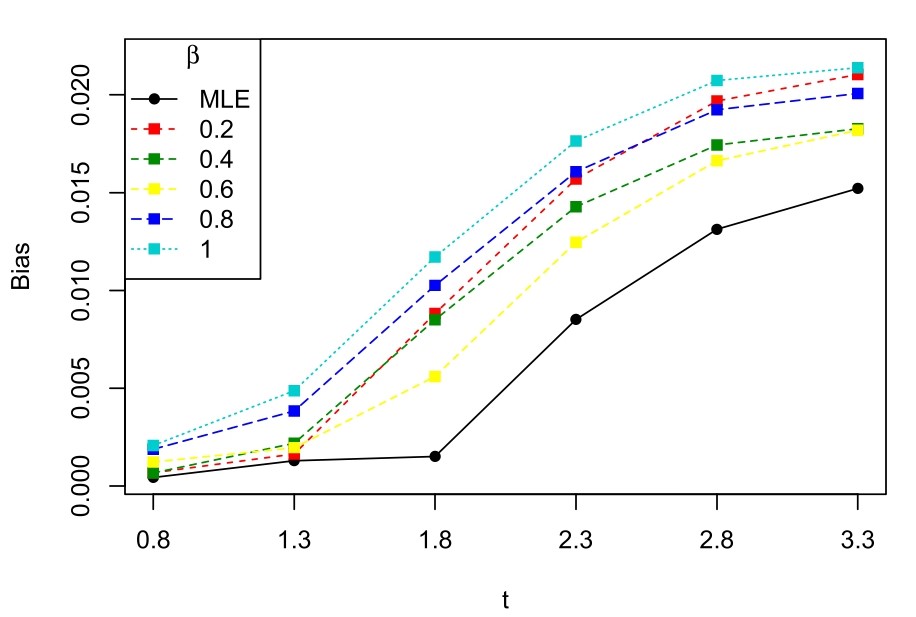}}
\subfloat[Group 3]{\includegraphics[width =0.5\textwidth]{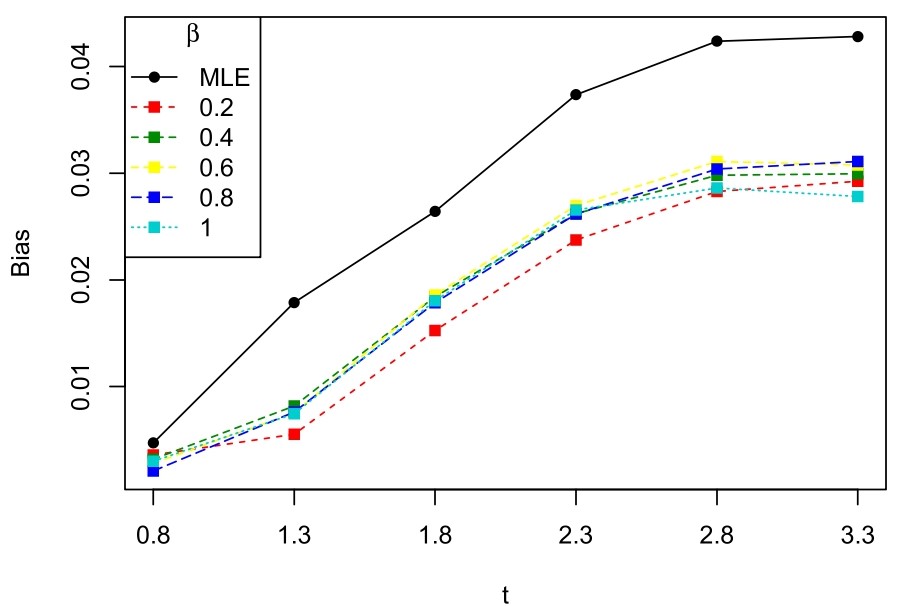}}
\end{center}
\caption{Bias of reliability estimates in pure (left) and contaminated (right) data set.}
\label{fig2}
\end{figure}

\noindent\textcolor{black}{In Tables \eqref{tab4}-\eqref{tab7}, bias and mean-square-error (MSE) of MLE and WMDPDE, as well as estimated reliability at time point $t=2.5$ for pure data and contamination settings, are provided.  It can be seen that the bias and MSE of MLE are lower than those of WMDPDE in the context of pure data.  Thus, MLE beats WMDPDE in the pure data scheme.  But in case of contamination in data, bias and MSE of WMDPDE are lower than those of MLE.  The increase of bias of WMDPDE from pure data to contaminated data is lesser than that of MLE.  This indicates that WMDPDE is a robust estimator without compromising efficiency much. 
 The graphical representation of bias of reliability estimates in pure data and contaminated data settings at different time points is given in Figure \ref{fig2}.  It can be visually observed that in the pure data scheme, MLE performs better than WMDPDE and in the contaminated data setting WMDPDE outperforms MLE in terms of bias.  Due to the superiority of WMDPDE numerically proven here, these estimators are the preferable choice over MLEs when data is contaminated.}

\begin{table}[htb!]
\color{black}
\caption{Bias of Estimates of parameters in pure and contaminated data setting}
\begin{center}
{\scalebox{0.88}{
\begin{tabular}{lcccccccc}
\hline%
& \multicolumn{4}{@{}c@{}}{\textbf{Pure Data}}
& \multicolumn{4}{@{}c@{}}{\textbf{Contaminated Data}}\\\cmidrule(lr){2-5}\cmidrule(lr){6-9}
&$\mathbf{a_1}$&$\mathbf{a_2}$&$\mathbf{b_1}$&$\mathbf{b_2}$&$\mathbf{a_1}$&$\mathbf{a_2}$&$\mathbf{b_1}$&$\mathbf{b_2}$\\
\hline
$\bm{\theta=\theta}_1$\\
$\textbf{MLE}$  &-0.0073  &-0.0231  &-0.0245  & -0.0010 &-0.0390  & -0.0878  &-0.0720 & -0.0639\\
	$\bm{\beta{=}0.2}$  & -0.0100 & -0.0303 &-0.0362 &-0.0020 & -0.0230  & -0.0561  &-0.0621 &-0.0441 \\
	$\bm{\beta{=}0.4}$ & -0.0137 &-0.0379 & -0.0402 &-0.0055 & -0.0249 & -0.0598  &-0.0646  & -0.0492\\
		$\bm{\beta{=}0.6}$ & -0.0162 & -0.0425 & -0.0432& -0.0069 & -0.0235 & -0.0570  &-0.0639 &-0.0479\\
		$\bm{\beta{=}0.8}$& -0.0139 & -0.0378 & -0.0414 & -0.0039 & -0.0211 & -0.0523  &-0.0631 &-0.0462 \\
		$\bm{\beta{=}1.0}$ & -0.0140 & -0.0381 &-0.0448 & -0.0099 &-0.0194  & -0.0489  &-0.0627 &-0.0454   \\
  \hline
$\bm{\theta=\theta}_2$ \\
$\textbf{MLE}$  & -0.0061 &0.0029 &0.0081 &-0.0419  &  -0.0221 & 0.0398  & 0.0278  & -0.0582\\
	$\bm{\beta{=}0.2}$  &-0.0147 & 0.0215 & 0.0152 &-0.0462  &-0.0104  & 0.0251 &0.0159 &-0.0472 \\
	$\bm{\beta{=}0.4}$ & -0.0170 & 0.0258 & 0.0145& -0.0495&-0.0177  &0.0245   &0.0184 &-0.0407 \\
		$\bm{\beta{=}0.6}$ & -0.0209 &0.0181 &0.0213  & -0.0354 &-0.0266  &  0.0201 & 0.0235&-0.0311\\
		$\bm{\beta{=}0.8}$& -0.0229 & 0.0140  &0.0248 &-0.0416  &-0.0241  & 0.0196  & 0.0262&-0.0262 \\
		$\bm{\beta{=}1.0}$ & -0.0221 &0.0156 &0.0227 &-0.0242  & -0.0229 & 0.0141  &0.0279 & -0.0231  \\
\hline
$\bm{\theta=\theta}_3$ \\
$\textbf{MLE}$  & -0.0010 & -0.0140 &0.0301 &-0.0130  &-0.0171  & -0.0634  &0.0717 &0.0622 \\
	$\bm{\beta{=}0.2}$  &0.0047 & -0.0204 &0.0367&-0.0060 & 0.0047 & -0.0203  &0.0490 &0.0179 \\
	$\bm{\beta{=}0.4}$ &0.0019 &-0.0260 &0.0307 &-0.0177  &0.0023  & -0.0124  &0.0499 &0.0199 \\
		$\bm{\beta{=}0.6}$ & 0.0031 & -0.0236 &0.0330 & -0.0133 & 0.0077 & -0.0145  &0.0482 & 0.0163\\
		$\bm{\beta{=}0.8}$& 0.0027 & -0.0184 &0.0371 &-0.0110  & 0.0034 & -0.0101  &0.0452 &0.0103 \\
		$\bm{\beta{=}1.0}$ & 0.0066 & -0.0167 &0.0327  & -0.0140 & 0.0073 & -0.0113  &0.0447 & 0.0113  \\
\hline
\end{tabular}}}
\end{center}
\label{tab4}
\end{table}

\begin{table}[htb!]
\color{black}
\caption{MSE of Estimates of parameters in pure and contaminated data setting}
\begin{center}
{\scalebox{0.88}{
\begin{tabular}{lcccccccc}
\hline%
& \multicolumn{4}{@{}c@{}}{\textbf{Pure Data}}
& \multicolumn{4}{@{}c@{}}{\textbf{Contaminated Data}}\\\cmidrule(lr){2-5}\cmidrule(lr){6-9}
&$\mathbf{a_1}$&$\mathbf{a_2}$&$\mathbf{b_1}$&$\mathbf{b_2}$&$\mathbf{a_1}$&$\mathbf{a_2}$&$\mathbf{b_1}$&$\mathbf{b_2}$\\
\hline
$\bm{\theta=\theta}_1$ \\
$\textbf{MLE}$  & 0.0080 &0.0345  &0.0164  &0.0721  &  0.0345 &0.1383  & 0.0456 &0.1626 \\
	$\bm{\beta{=}0.2}$  &0.0087  &0.0351  & 0.0249 &0.0881 &0.0122  &0.0498  & 0.0222   & 0.0734  \\
	$\bm{\beta{=}0.4}$ &0.0165  & 0.0714 &0.0174  &0.0930 & 0.0110  &0.0451 &0.0296  & 0.1024 \\
		$\bm{\beta{=}0.6}$ & 0.0174 & 0.0704 &0.0265  & 0.0972 &0.0106  &0.0433 &0.0259  & 0.0882 \\
		$\bm{\beta{=}0.8}$&0.0146  &0.0592  & 0.0210 &0.0639 &0.0094  &0.0385   & 0.0204   & 0.0667  \\
		$\bm{\beta{=}1.0}$ &0.0094  & 0.0382 & 0.0181 & 0.0635 & 0.0033  &0.0141 &0.0221 &  0.0738  \\
\hline
$\bm{\theta=\theta}_2$ \\
$\textbf{MLE}$  & 0.0600 & 0.3946 & 0.0287 & 0.1177 & 0.0717 &0.3467  & 0.0229  & 0.1815 \\
	$\bm{\beta{=}0.2}$  & 0.0476 &0.1829  &0.0251  &0.0975  &  0.0525     & 0.1919    & 0.0321    &0.1236   \\
	$\bm{\beta{=}0.4}$& 0.0372 & 0.1482 & 0.0306 & 0.1097  &0.0212 & 0.0842  & 0.0217  & 0.0841 \\
		$\bm{\beta{=}0.6}$ &0.0231 & 0.0908 &0.0213  & 0.0818  &0.0406  &0.1672  & 0.0170   & 0.0641 \\
		$\bm{\beta{=}0.8}$&0.0321  &0.1264  &0.0167    &0.0693  & 0.0211    & 0.0798  & 0.0187  & 0.0708 \\
		$\bm{\beta{=}1.0}$& 0.0136 & 0.0529 & 0.0152   &  0.0523 & 0.0121  & 0.0467  & 0.0134  &  0.0494   \\
\hline
$\bm{\theta=\theta}_3$ \\
$\textbf{MLE}$& 0.0170   & 0.0472  & 0.0291    & 0.0828 & 0.0392 &0.1560   &0.0496   &0.1742 \\
	$\bm{\beta{=}0.2}$& 0.0050  & 0.0201 & 0.0140   & 0.0790 & 0.0106   & 0.0423  & 0.0193  & 0.0649 \\
	$\bm{\beta{=}0.4}$& 0.0088  & 0.0360  & 0.0173 & 0.0637& 0.0175 & 0.0498   & 0.0199  & 0.0681\\
		$\bm{\beta{=}0.6}$& 0.0099   & 0.0400  & 0.0177 &  0.0650   & 0.0084  &0.0334   & 0.0167  & 0.0564  \\
		$\bm{\beta{=}0.8}$& 0.0079   & 0.0208  & 0.0121  &  0.0485 & 0.0146 & 0.0321 & 0.0115 & 0.0368 \\
		$\bm{\beta{=}1.0}$& 0.0034  & 0.0138 & 0.0098  & 0.0342  &0.0105  & 0.0269 & 0.0136  & 0.0493 \\
\hline
\end{tabular}}}
\end{center}
\label{tab5}
\end{table}

\begin{table}[htb!]
\color{black}
\caption{Bias of Reliability estimates of parameters in pure and contaminated data setting}
\begin{center}
{\scalebox{0.88}{
\begin{tabular}{lcccccc}
\hline%
& \multicolumn{3}{@{}c@{}}{\textbf{Pure Data}}
& \multicolumn{3}{@{}c@{}}{\textbf{Contaminated Data}}\\\cmidrule(lr){2-4}\cmidrule(lr){5-7}
&\textbf{Group 1}&\textbf{Group 2}&\textbf{Group 3}&\textbf{Group 1}&\textbf{Group 2}&\textbf{Group 3}\\
\hline
$\bm{\theta=\theta}_1$ \\ 
$\textbf{MLE}$  & 0.0057  &0.0099  &  0.0107  & 0.0219  & 0.0338 & 0.0423  \\
	$\bm{\beta{=}0.2}$  & 0.0087  & 0.0141 & 0.0181  & 0.0143  & 0.0221 & 0.0296   \\
	$\bm{\beta{=}0.4}$ & 0.0079  & 0.0125 & 0.0165  & 0.0135  & 0.0215 &  0.0287  \\
		$\bm{\beta{=}0.6}$ & 0.0089  & 0.0139 & 0.0183  &  0.0126 & 0.0203 & 0.0278  \\
		$\bm{\beta{=}0.8}$& 0.0079  & 0.0133 & 0.0177  & 0.0128  & 0.0199 & 0.0264   \\
		$\bm{\beta{=}1.0}$ & 0.0086  &0.0131  & 0.0174  & 0.0138  & 0.0221 & 0.0303   \\
\hline
$\bm{\theta=\theta}_2$ \\
$\textbf{MLE}$  & 0.0017  &0.0038  & 0.0057  & 0.0067  & 0.0095 & 0.0114  \\
	$\bm{\beta{=}0.2}$  &  0.0054 &0.0075  & 0.0092  &  0.0047 &0.0061 &  0.0075  \\
	$\bm{\beta{=}0.4}$& 0.0049  & 0.0062 & 0.0075  &  0.0045 & 0.0060 & 0.0075  \\
		$\bm{\beta{=}0.6}$ & 0.0059  & 0.0076 & 0.0093  & 0.0050  & 0.0062 & 0.0076   \\
		$\bm{\beta{=}0.8}$& 0.0054  & 0.0069 & 0.0083  &   0.0058 & 0.0071 &  0.0084  \\
		$\bm{\beta{=}1.0}$& 0.0045  & 0.0059 & 0.0073  & 0.0040  & 0.0049 &  0.0060   \\
\hline
$\bm{\theta=\theta}_3$ \\
$\textbf{MLE}$& 0.0049  & 0.0092 & 0.0135  & 0.0076  & 0.0188 & 0.0217  \\
	$\bm{\beta{=}0.2}$& 0.0027  & 0.0060 & 0.0093  & 0.0020  & 0.0063 & 0.0087 \\
	$\bm{\beta{=}0.4}$& 0.0059  & 0.0102 & 0.0133  &  0.0021 & 0.0048 & 0.0064 \\
		$\bm{\beta{=}0.6}$& 0.0072  & 0.0136  & 0.0174  & 0.0018  & 0.0039 & 0.0051  \\
		$\bm{\beta{=}0.8}$& 0.0041  & 0.0081 & 0.0113  &  -0.0010  & 0.0057 & 0.0054 \\
		$\bm{\beta{=}1.0}$& 0.0025  & 0.0049 & 0.0075  &  -0.0022 & -0.0020 & -0.0014  \\
\hline
\end{tabular}}}
\end{center}
\label{tab6}
\end{table}

\begin{table}[htb!]
\color{black}
\caption{MSE of Reliability estimates of parameters in pure and contaminated data setting}
\begin{center}
{\scalebox{0.88}{
\begin{tabular}{lcccccc}
\hline%
& \multicolumn{3}{@{}c@{}}{\textbf{Pure Data}}
& \multicolumn{3}{@{}c@{}}{\textbf{Contaminated Data}}\\\cmidrule(lr){2-4}\cmidrule(lr){5-7}
&\textbf{Group 1}&\textbf{Group 2}&\textbf{Group 3}&\textbf{Group 1}&\textbf{Group 2}&\textbf{Group 3}\\
\hline
$\bm{\theta=\theta}_1$ \\ 
$\textbf{MLE}$ & 0.0015  & 0.0039  & 0.0064  & 0.0053  & 0.0114  & 0.0147  \\
	$\bm{\beta{=}0.2}$ & 0.0024 & 0.0057 & 0.0082 & 0.0023  & 0.0044  & 0.0074     \\
	$\bm{\beta{=}0.4}$ & 0.0021  & 0.0045  & 0.0071&  0.0023  & 0.0053 & 0.0083 \\
		$\bm{\beta{=}0.6}$ &0.0022 & 0.0045 & 0.0070 & 0.0017  & 0.0041  & 0.0069 \\
		$\bm{\beta{=}0.8}$  & 0.0018  & 0.0046  & 0.0072& 0.0023 & 0.0048 & 0.0074  \\
		$\bm{\beta{=}1.0}$& 0.0022  & 0.0040 & 0.0066  & 0.0015 & 0.0036 & 0.0064  \\
\hline
$\bm{\theta=\theta}_2$ \\
$\textbf{MLE}$  & 0.0028  & 0.0043 & 0.0048 & 0.0048 & 0.0070 &0.0078 \\
	$\bm{\beta{=}0.2}$  & 0.0033 & 0.0048 & 0.0053& 0.0035  & 0.0043  & 0.0047  \\
	$\bm{\beta{=}0.4}$ & 0.0033 & 0.0041 & 0.0045   & 0.0032   & 0.0043  & 0.0049 \\
		$\bm{\beta{=}0.6}$  & 0.0033 & 0.0044 & 0.0050 & 0.0028   &0.0035  & 0.0039  \\
		$\bm{\beta{=}0.8}$ & 0.0026  & 0.0033  & 0.0036  & 0.0031   & 0.0038  & 0.0041\\
		$\bm{\beta{=}1.0}$ & 0.0023   & 0.0031  & 0.0036 & 0.0024   & 0.0032  & 0.0036 \\
\hline
$\bm{\theta=\theta}_3$ \\
$\textbf{MLE}$  & 0.0012  & 0.0032  & 0.0052  & 0.0042 & 0.0171  & 0.0205 \\
	$\bm{\beta{=}0.2}$& 0.0009 & 0.0028 & 0.0048 & 0.0018  &0.0064  & 0.0087\\
	$\bm{\beta{=}0.4}$& 0.0021 & 0.0053  & 0.0067& 0.0019 & 0.0051  & 0.0062 \\
		$\bm{\beta{=}0.6}$ & 0.0022 & 0.0076 & 0.0092& 0.0019 & 0.0046  &0.0056 \\
		$\bm{\beta{=}0.8}$ & 0.0013 & 0.0039  & 0.0054  & 0.0010 & 0.0034  & 0.0058\\
		$\bm{\beta{=}1.0}$ & 0.0010 & 0.0023 & 0.0035  & 0.0007 &0.0019  & 0.0029  \\
\hline
\end{tabular}}}
\end{center}
\label{tab7}
\end{table}

\section{Optimum Inspection Time}\label{sec5}

\noindent One-shot devices stay in a torpid state until used.  Hence the behaviour of such devices is revealed only when these are put to the test.  The aim is to find a set of inspection times which will \textcolor{black}{increase the} precision \textcolor{black}{of the estimator} and reduce the experimental cost due to the destruction of experimental units.  Therefore, the cost function \textcolor{black}{can be} defined as follows.
\begin{equation}
	\label{cost}
	Cost=C_1\times \vert V\vert +C_2\times E(\sum_{i=1}^{I}n_i)
\end{equation}
\textcolor{black}{Here} $\vert V\vert$ is the determinant value of the covariance matrix based on the asymptotic distribution of WMDPDE, $E(\sum_{i=1}^{I}n_i)$ is the expected total number of failures from all I different groups.  Cost $C_1$ is imposed on $\vert V\vert $, which reflects the precision cost of the estimator and $C_2$ is the cost per failure of experimental units.  Given the values of group sizes, the optimal inspection times will minimize the cost function defined in \eqref{cost}.  This cost function brings a trade-off between the precision of the estimation and the experimental cost.

\subsection{\textbf{Search Algorithm}}
\noindent Here, the objective is to find \textcolor{black}{the set of} inspection times which will minimize the above-defined cost. \textcolor{black}{ We apply} population-based heuristic Natural optimization method, namely Genetic Algorithm (GA), which was developed by Holland \cite{jh1975adaptation} and popularized by Goldberg \cite{golberg1989genetic}.  GA is an optimization method that produces generations of the population through selection, crossover and mutation.  It applies to discrete as well as continuous variables and can deal with a large number of variables.  It can perform optimization with extremely complex cost surfaces and is capable of discovering global optimums avoiding the trap of local optima.  GA is inherently parallel, modifiable, easily distributed and adaptable to different problems \cite{ou2014using}.  GA works well with numerically generated data, experimental data, or analytical functions \cite{haupt2004practical}.  Some of the applications of GA can be seen in Babaei and Sheidaii \cite{babaei2013optimal}, Liou et al. \cite{liou2013genetic}, Soolaki et al. \cite{soolaki2012new}, Lopez et al. \cite{lopez2019behavior} and references therein. \\

\noindent To search the optimal inspection times through GA, the strategy of keeping the best or the elite observations of the current population in the new population is followed; readers may see Chakraborty and Chaudhuri \cite{chakraborty2003use} for detailed discussion.  Thus traits of the observations with the best cost, i.e. minimum cost, are retained from generation to generation.  The \textcolor{black}{steps} of GA using elitism is as follows.
\begin{enumerate}[label=\roman*)]
	\item Set the initial population size $N_{pop}$ where each element in the population is a set of inspection times with dimension $N_{var}=I$.
	
	\item  Obtain the cost for each set of inspection times and select the parents for pairing.	
	
	\item The parents are selected using a specified procedure based on cost.  Among the various selection methods, the random rank selection procedure and tournament selection procedure are used, which are explained below.
	
	\item Each set of parents would produce two offspring.  Odd numbered element is chosen as the mother, and an even-numbered element is chosen as the father.  The offsprings \textcolor{black}{are produced as:}
	$$\text{Offspring 1}= b\times \text{mother}+(1-b)\times \text{father}$$
	$$	\text{Offspring 2}= (1-b)\times \text{mother}+b\times \text{father}$$
	where $b$ is a random \textcolor{black}{uniform} number.  Thus a new population is formed containing the parents and offspring.

	\item Now, the population is mutated with the mutation rate $mr$.  So the number of mutations is $mn=mr\times N_{pop}\times N_{var}$.  The values in randomly selected $mn$ positions out of $N_{pop}\times N_{var}$ positions are replaced with the randomly generated new values.  Hence a mutated population is obtained, which would function as the initial population for the next generation.
	
	\item The process is repeated up to, say, B generations.  For each generation, a local minimum is obtained, \textcolor{black}{i.e.} the set of inspection times with the least cost.  Among these B local minima, the global minimum is obtained: the set of inspection times with minimum cost \textcolor{black}{based on all the generated populations.}  This global minimum \textcolor{black}{provides} the set of optimum inspection times.
 
	\subsubsection{Random Rank Selection Procedure}
	\begin{itemize}
		\item \textcolor{black}{Rank} the observations concerning the cost in increasing order.
		\item Define $N_{keep}=0.5\times N_{pop}$.
		\item Keep \textcolor{black}{first} $N_{keep}$ observations and discard the rest.
		\item Use a permutation of ranks and rearrange the $N_{keep}$ size data concerning this permutation.
		\item The resultant observations are chosen as the parents who are to be paired.
	\end{itemize}
	\subsubsection{Tournament Selection Procedure}
	\begin{itemize}
		\item  Randomly pick a pair of observations such that this pair is not selected again.
		\item The observation having lesser cost in the pair is kept as a parent, and the other is discarded.  Though the same pair can not be chosen again, the discarded observation or the chosen observation in the pair can be chosen again for pairing.
		\item Repeat this process up to $N_{keep}$ times.
		\item The resultant observations are chosen as the parents who are to be paired.
	\end{itemize}
\end{enumerate}
\begin{table}[htb!]
\caption{Model Parameters}
\begin{center}
{\begin{tabular}{lllll}
\hline
\textbf{S.No.}& $\;\mathbf{a_1}\;$ & $\;\mathbf{a_2}\;$ & $\;\mathbf{b_1}\;$ & $\;\mathbf{b_2}\;$\\
\hline
	$\bm{\theta}_1$ & $0.2$ & $-0.6$&$-0.2$ &$0.4$ \\
	$\bm{\theta}_2$& $-0.3$& $-0.1$ & $-0.2$ & $0.1$\\
\hline
\end{tabular}}
\end{center}
\label{tab8}
\end{table}	
\noindent Here, a numerical experiment is conducted to derive optimal inspection times implementing GA under the following setups.  The chosen model parameters are given in Table \eqref{tab8} with
	tuning parameter $\beta=0.1$, and the layout is given in  Table \eqref{tab9}. 
	Costs are set as  $C_1=C_2=0.5$ where
	$N_{pop}=12,$ $N_{var}=5.$  GA continues till $B=250$ generations. In the crossover,
	the random number $b$ follows $U(0,1),$ where in the mutation stage, the mutation rate is set as $mr=0.20,$ and the value in the selected position is replaced by a new random number from $U(0,10)$.\\ 
\begin{table}[htb!]
\caption{Layout for the implementation of GA}
\begin{center}
{\begin{tabular}{lcccc}
\hline%
& &  & \multicolumn{2}{l}{Covariates} \\\cline{4-5}%
Groups  & Devices & Failures & Stress 1 & Stress 2 \\
\hline
1 &  $8$&$n_1$ & $20$ & $40$ \\
2 &  $12$&$n_2$ &$30$ & $60$\\
3&$16$ &$n_3$ &$40$ & $80$\\
4&$20$ &$n_4$ &$30$ & $40$\\
5&$24$ &$n_5$ &$20$ & $50$\\
\hline
\end{tabular}}
\end{center}
\label{tab9}
\end{table}

\noindent The Optimum inspection times $\tau_1,\tau_2,\tau_3,\tau_4,\tau_5$ by the random rank selection and tournament selection procedure are given in the Table \eqref{tab10}.  Though the two selection procedures yield very different sets of inspection times yet the resultant costs for both of them are very similar.  Thus these two selection procedures can be used alternatively for Genetic Algorithm.
\begin{table}[htb!]
\caption{Optimum Inspection Times}
\begin{center}
{\scalebox{0.88}{
\begin{tabular}{lllllll}
\hline
	\textbf{Selection Method}&$\mathbf{\tau_1}$ & $\mathbf{\tau_2}$ & $\mathbf{\tau_3}$& $\mathbf{\tau_4}$ & $\mathbf{\tau_5}$& \textbf{Cost} \\
\hline
$\bm{\theta}=\bm{\theta}_1$ & & & & & & \\
Random rank selection &6.13382 &0.98725 & 2.69014& 8.90894&4.05132 &2.46078\\
Tournament selection &8.34024 &1.56001 & 7.02301& 0.98418&0.21274 &2.32247\\
\hline
$\bm{\theta}=\bm{\theta}_2$ & & & & & & \\
Random rank selection &9.18581 &2.63731 &9.62269 &0.05032 &4.53493 &1.80780\\
Tournament selection &4.29600 & 9.97924&0.67811& 8.40763&0.02557 &1.95393\\
\hline
\end{tabular}}}
\end{center}
\label{tab10}
\end{table}	

\section{Application to Real Dataset}\label{sec6}
\noindent \textcolor{black}{Actual data is adopted from the database named ``Incidence$\cdot$SEER Research Data, 8 Registries, Nov 2021 Sub (1975-2019)"\cite{seer} which is recorded by National Cancer Institute (Surveillance, Epidemiology, and End Results Program (SEER)).  We have extracted data from patients between the periods 2016-2019 who are diagnosed with gallbladder cancer.  The patients of three age groups (40-49), (50-59), and (60-69) are considered for the study.  Their median age at diagnosis ($x_{i1}$) and the size of the tumour ($x_{i2}$) are taken as two stress factors/covariates.  The median age is divided by 10 for the sake of computational ease.  The size of the tumour is indicated below,
$$
x_{i2}=\begin{cases}
			1, & \text{if tumor size}\leq 50\,mm\\
            2, & \text{otherwise}
		 \end{cases}
$$  
Those patients are observed over a number of months, and the number of failures was reported within follow-up times where death due to gallbladder cancer indicates failure.  It is also observed from the data that, there are some patients who died within a month of diagnosis of disease and could not contribute to the study.  Those patients can be considered outliers.  The description of data is given in Table \eqref{tab11}.  }\\
\begin{table}[htb!]
\color{black}
\caption{Layout of Dataset extracted from \cite{seer}}
\begin{center}
{\begin{tabular}{lccccc}
\hline%
& & & & \multicolumn{2}{c}{\textbf{Covariates}}\\\cline{5-6}%
\textbf{Groups} & \textbf{Observation} & \textbf{Patients} & \textbf{Death} & \textbf{Median} & \textbf{Tumor} \\
& \textbf{months} & \textbf{count} & \textbf{count} & \textbf{age} & \textbf{size} \\
\hline
1 & 12 & 17 & 12 & 46.40 & 1\\
2 & 12 & 12 & 10 & 46.40 & 2 \\
3 & 12 & 41 & 32 & 56.02 & 1\\
4 & 12 & 48 & 42 & 56.02 & 2 \\
5 & 12 & 95 & 61 & 64.96 & 1 \\
6 & 12 & 74 & 64 & 64.96 & 2 \\
\hline
\end{tabular}}
\end{center}
\label{tab11}
\end{table}

\noindent\textcolor{black}{To check if a lifetime distribution can be fitted to the data, bootstrap testing is conducted based on the distance-based test statistic proposed by Balakrishnan and Ling \cite{balakrishnan2012multiple}.  The test statistic is given as,
$$
D=max_i\vert n_i-e_i\vert
$$
where $n_i$ is the observed failure, and $e_i$ is the expected failure for $i=1,\ldots,6$.  The MLE is used to estimate the expected failures, $e_i=k_i\times F(\tau_i;x_{ij},\bm{\theta})$.  The bootstrap (BT) method is utilized to approximate the p-value of the test based on 1000 generated samples.  The test statistic and corresponding p-value for Weibull, Gamma and Logistic-Exponential distribution are given in Table \eqref{tab12}.  The approximate p-value from this table suggests the suitability of the Logistic-Exponential model to the given data.  Therefore, we assume that the lifetime of gallbladder cancer patients follows the LE distribution.} \\

\begin{figure}
\color{black}
\begin{center}
\subfloat[Group 1]{\includegraphics[width =0.5\textwidth]{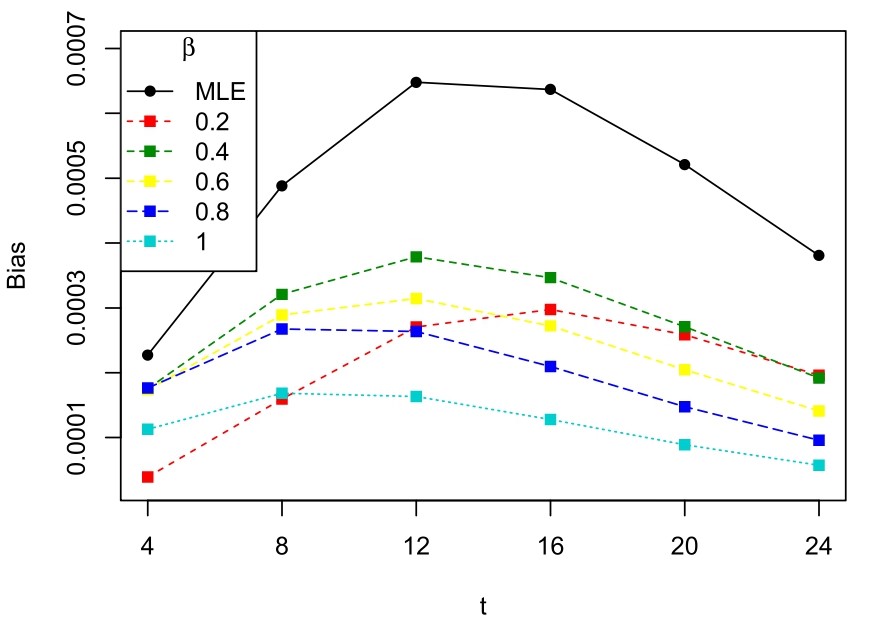}} 
\subfloat[Group 2]{\includegraphics[width =0.5\textwidth]{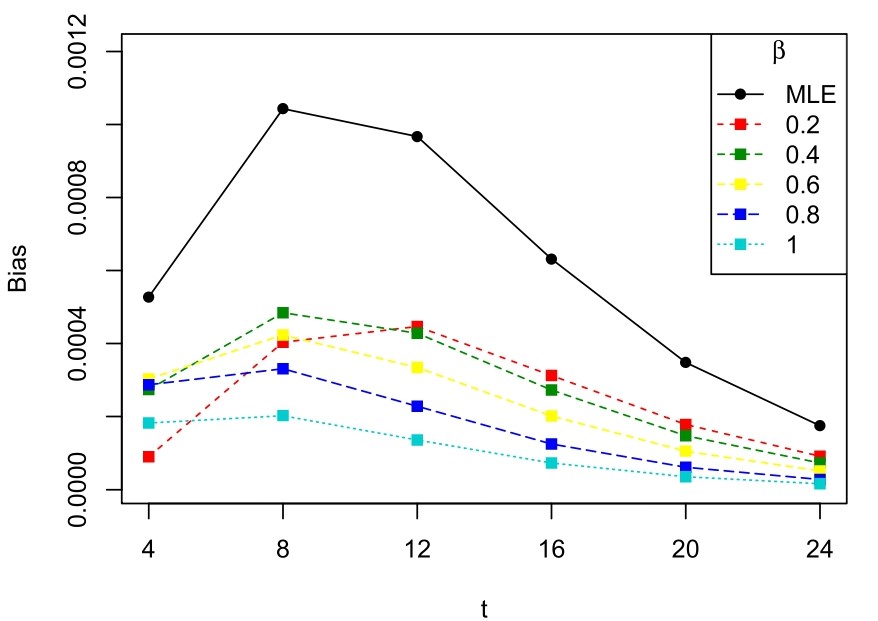}}\\
\subfloat[Group 3]{\includegraphics[width =0.5\textwidth]{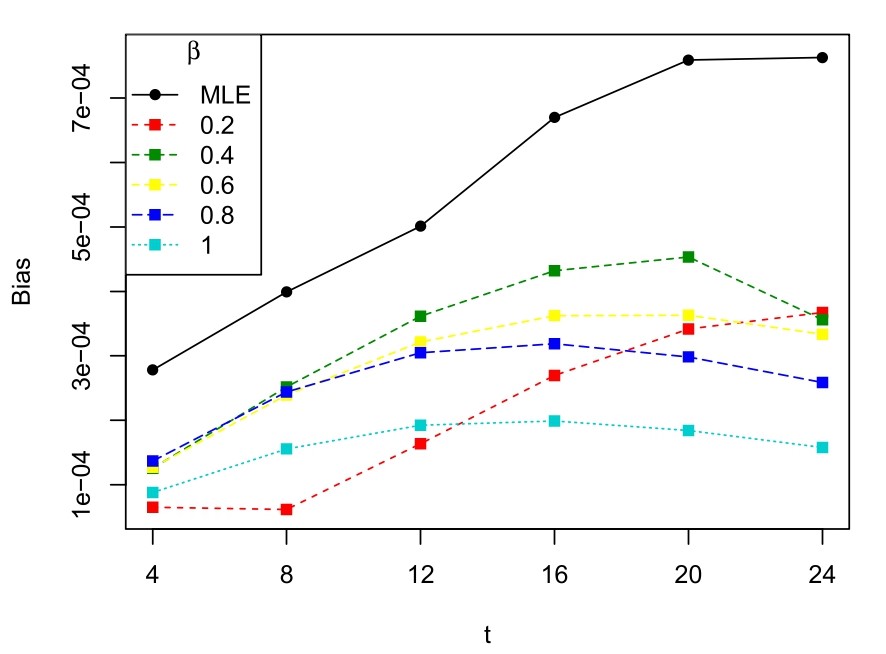}}
\subfloat[Group 4]{\includegraphics[width =0.5\textwidth]{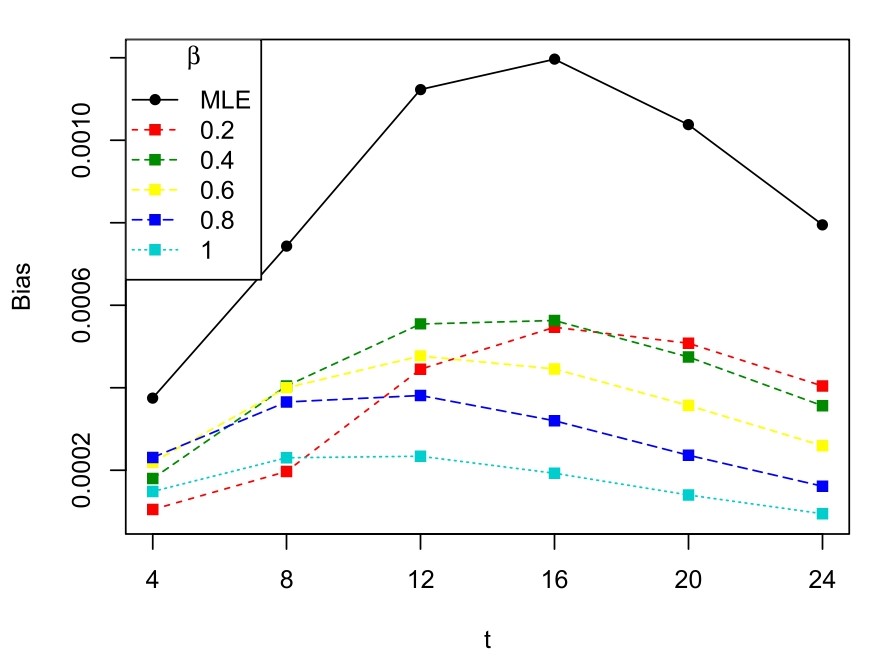}} \\
\subfloat[Group 5]{\includegraphics[width =0.5\textwidth]{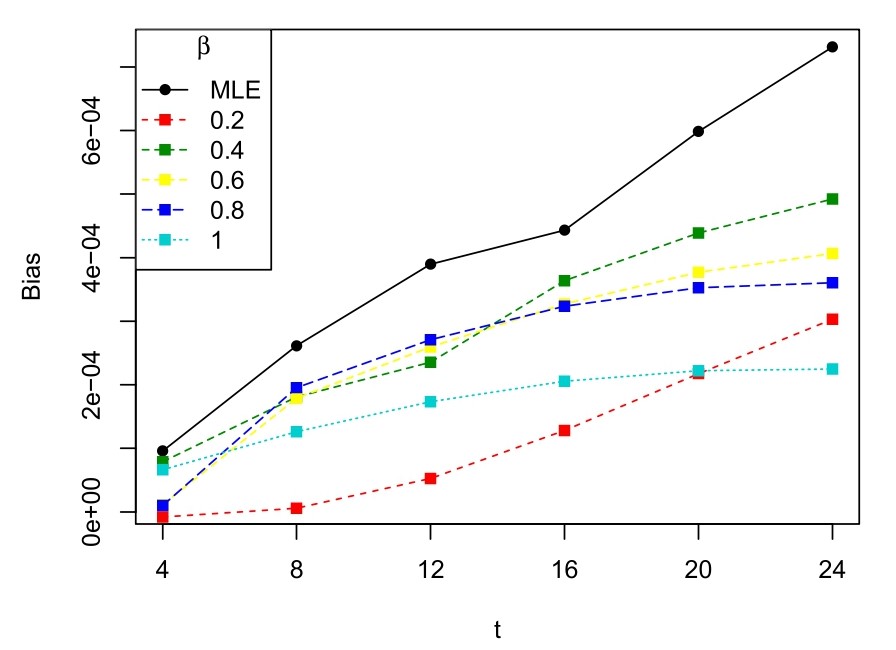}}
\subfloat[Group 6]{\includegraphics[width =0.5\textwidth]{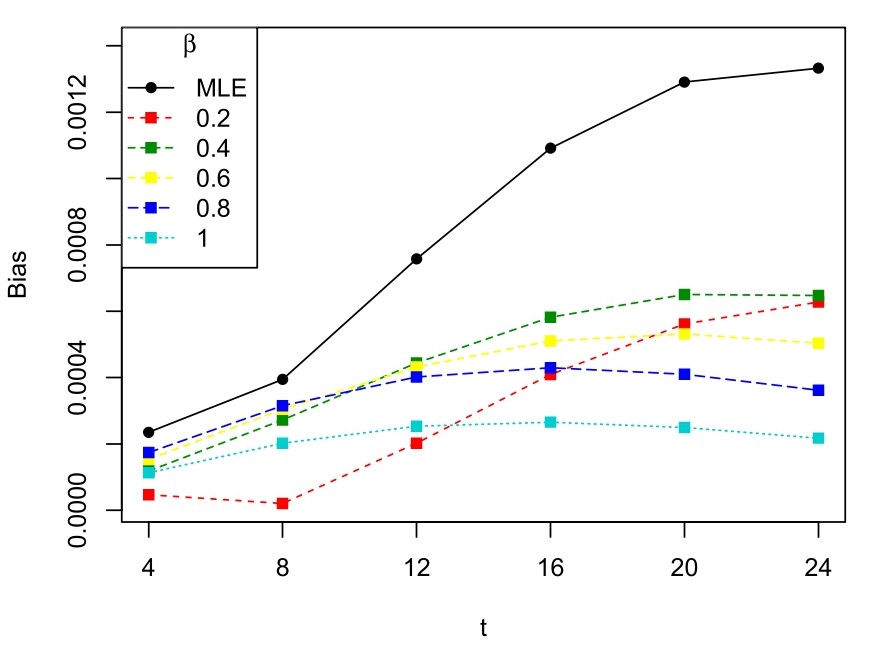}}
\end{center}
\caption{Bias of reliability estimates for the real dataset.}
\label{fig3}
\end{figure}

\noindent \textcolor{black}{MLE and WMDPDE of the model parameters are obtained, and bootstrap estimates of bias and root-mean-square-error generated from 1000 bootstrap samples are computed.  The coordinate descent algorithm's initial parameter value is set as (-4.46, 0.08, -0.21, 0.34), which is chosen through an extensive grid search process.  The outcomes of the MLE and WMDPDE are reported in Tables \eqref{tab13} and \eqref{tab14}.  The graphical representation of the bias of reliability estimates at different time points is given in Figure \ref{fig3}.  It is observed that WMDPDE have lower bias and RMSE than those of MLE.  Since the data is contaminated, the superiority of WMDPDE in this situation validates the objective of our study.}
\begin{table}[htb!]
 \color{black}
\caption{Test statistic and corresponding p-value calculation}
\begin{center}
{\begin{tabular}{lcl}
\hline
\textbf{Lifetime Distribution}& \textbf{Test Statistic} & \textbf{p-value}\\
\hline
Weibull & 12.6873 & 0.004 \\
Gamma & 33.5532 & 0.005 \\
Logistic Exponential & 06.0709 & 0.192 \\
\hline
\end{tabular}}
\end{center}
\label{tab12}
\end{table}	

\begin{table}[htb!]
\color{black}
\caption{Estimates of parameters for real dataset}
\begin{center}
{\scalebox{0.88}{
\begin{tabular}{lcccccc}
\hline%
&\textbf{MLE}&$\bm{\beta=0.2}$&$\bm{\beta=0.4}$&$\bm{\beta=0.6}$&$\bm{\beta=0.8}$&$\bm{\beta=1.0}$\\
\hline
$\hat{\bm{a}}_1$  & -4.463718 & -4.462089  & -4.461218  &-4.460754  & -4.460504 &-4.460369 \\
	\textbf{Bias}  & 0.002955 & 0.002513 & -0.000923 & -0.000666 &0.000471  & -0.000070  \\
	\textbf{RMSE}& 0.059291 &0.053992  &0.044467  & 0.029429 & 0.022231 & 0.012952 \\
\hline
$\hat{\bm{a}}_2$  & 0.080093  & 0.080022 & 0.080026 & 0.080026 & 0.080023 & 0.080020 \\
	\textbf{Bias}  & 0.000089 & 0.000058 &-0.000023  & -0.000014 &  0.000007& -0.000001  \\
	\textbf{RMSE} & 0.001739 & 0.001613 &  0.001240 & 0.000816 & 0.000608 & 0.000347 \\
\hline
$\hat{\bm{b}}_1$  &-0.210451  & -0.210285 & -0.210189 & -0.210132 &  -0.210098 & -0.210077 \\
	\textbf{Bias}  & 0.000333  &0.000176  & -0.000014 &  -0.000048 & 0.000032 & -0.000005  \\
	\textbf{RMSE} & 0.004724 & 0.004219  & 0.003337 & 0.002143 & 0.001630 & 0.000902 \\
\hline
$\hat{\bm{b}}_2$  & 0.339126 &  0.339590 &  0.339825 & 0.339939 & 0.339991 & 0.340012 \\
	\textbf{Bias} &0.000906  & 0.000457 &-0.000069  &-0.000102  & 0.000063 &  -0.000015 \\
	\textbf{RMSE} & 0.014378 & 0.012907 & 0.009650 &0.006169  & 0.004589 & 0.002562 \\
 \hline
\end{tabular}}}
\end{center}
\label{tab13}
\end{table}

\begin{table}[htb!]
\color{black}
\caption{Reliability estimates of parameters for real dataset}
\begin{center}
{\scalebox{0.88}{
\begin{tabular}{lcccccc}
\hline%
&\textbf{MLE}&$\bm{\beta=0.2}$&$\bm{\beta=0.4}$&$\bm{\beta=0.6}$&$\bm{\beta=0.8}$&$\bm{\beta=1.0}$\\
\hline
\textbf{Group 1} \\
\textbf{Bias}  & 0.000818  & 0.000049 &0.000341  & 0.000212 & 0.000008 & 0.000051  \\
\textbf{RMSE}&0.009974 & 0.008298 &  0.006667 & 0.004400 & 0.003277 & 0.001902 \\
\hline
\textbf{Group 2} \\
\textbf{Bias}  &0.000991  & 0.000299 & 0.000410 & 0.000221 & 0.000037 & 0.000042  \\
\textbf{RMSE}&0.008364  & 0.006788 &0.005318  & 0.003529 & 0.002575 &  0.001501 \\
\hline
\textbf{Group 3} \\
\textbf{Bias}  &0.000763  & -0.000118 &0.000309  & 0.000216 &-0.000014  & 0.000058  \\
\textbf{RMSE}&0.012092  & 0.010133 & 0.008180 & 0.005371 & 0.004052 & 0.002333 \\
\hline
\textbf{Group 4} \\
\textbf{Bias}  & 0.001228 & 0.0001184 &0.000508  & 0.000305 &0.000008  &  0.000064  \\
\textbf{RMSE}&0.015134  & 0.012468 & 0.009902 & 0.006535 & 0.004852 &  0.002809\\
 \hline
 \textbf{Group 5} \\
\textbf{Bias}  &0.000569 & -0.000213 & 0.000222 & 0.000174 & -0.000028 &  0.000050  \\
\textbf{RMSE}&0.011115  & 0.009351 & 0.007570 & 0.004949 & 0.003769 & 0.002156 \\
 \hline
 \textbf{Group 6} \\
\textbf{Bias}  & 0.000940 & -0.000202 &0.000379  & 0.000267 & -0.000038 &  0.000062 \\
\textbf{RMSE}& 0.017221 & 0.014293  & 0.011427 & 0.007481 & 0.005637 & 0.003233 \\
 \hline
\end{tabular}}}
\end{center}
\label{tab14}
\end{table}

\noindent \textcolor{black}{For robust WDPD-based testing, the null and alternative hypothesis is taken as follows.	$$H_0:\bm{\theta}=\bm{\theta}_0\qquad\text{against}\qquad H_1:\bm{\theta}\ne\bm{\theta}_0$$
	where, $\bm{\theta}_0= (-4.46, 0.08, -0.21, 0.34)^{'}$.  The values of WDPD-based test statistic are given in Table \eqref{tab15}.  The $rank(\nabla^2D^w_{\beta}(\bm{\theta}_0,\bm{\theta}_0)\Sigma_{\beta}(\bm{\theta}_0))=4$ and $\chi^2_{(4,0.95)}=0.711$.  From the values of the test statistic in Table \eqref{tab15}, it is seen that in all cases $\chi^2_{(4,0.95)}$ value is greater than the value of the observed test statistic.  Hence the null hypothesis fails to be rejected.}
\begin{table}[htb!]
 \color{black}
\caption{WDPD-based test statistic}
\begin{center}
{\begin{tabular}{cc}
\hline
\textbf{Tuning parameter}& \textbf{Test Statistic}\\
\hline
$\beta=0.2$ & 0.000998\\
 $\beta=0.4$ & 0.000887\\
$\beta=0.6$ & 0.000791\\
$\beta=0.8$ &  0.000712\\
$\beta=1.0$ & 0.000648\\
\hline
\end{tabular}}
\end{center}
\label{tab15}
\end{table}	
\section{Discussion and Conclusion}\label{sec7}
\noindent\textcolor{black}{This work proposed a robust method of estimation, the robust weighted minimum density power divergence estimators (WMDPDE), for point estimation of a Logistic-Exponential lifetime distribution based on one-shot device failure data. The derived estimators have been shown numerically to be more robust than the conventional Maximum Likelihood Estimation (MLE) when dealing with contaminated data. The asymptotic properties of the WMDPDEs and the density power divergence-based test statistic were provided.  Furthermore, a search for optimal inspection times was conducted using a Genetic Algorithm based on a cost function that balances the precision of the estimates and experimental cost. Finally, the proposed method was applied to a real data set where the Logistic-Exponential distribution was found to be a good fit. 
}\\

\noindent \textcolor{black}{In future we can incorporate several inspection times per group where failures are subject to multiple competing risks.  To conduct the experiment in a limited time and budget, we may apply progressive censoring in life-testing experiments of one-shot devices.} Works in those directions are in the pipeline, and we are optimistic about reporting those findings soon.
\textcolor{black}{
\section*{Acknowledgements}
\noindent The authors would like to thank two unknown reviewers and also the associate editor for many constructive suggestions which have helped to improve the paper significantly.}

\section*{Declarations of interest}
\noindent The authors declare no competing interests.

\section*{Funding}
\noindent This research did not receive any specific grant from funding agencies in the public, commercial, or not-for-profit sectors.

\textcolor{black}{
\section*{Data availability}
\noindent {\footnotesize Surveillance, Epidemiology, and End Results (SEER) Program (www.seer.cancer.gov) SEER*Stat Database: Incidence - SEER Research Data, 8 Registries, Nov 2021 Sub (1975-2019) - Linked To County Attributes - Time Dependent (1990-2019) Income/Rurality, 1969-2020 Counties, National Cancer Institute, DCCPS, Surveillance Research Program, released April 2022, based on the November 2021 submission.}}

\appendix
\textcolor{black}{
\section{Elements of Hessian Matrix for MLE}
\begin{flalign}
\frac{\partial^2\big(ln\,L(\bm{\theta})\big)}{\partial a_{j_1}\partial a_{j_2}}&=\sum_{i=1}^{I}\frac{\alpha_i x_{ij_1}x_{ij_2}\,ln\,(e^{\lambda_i\tau_i}-1)}{\{1+(e^{\lambda_i\tau_i}-1)^{\alpha_i}\}^2}\Big[k_i\alpha_i(e^{\lambda_i\tau_i}-1)^{\alpha_i}\,\,ln\,(e^{\lambda_i\tau_i}-1)\Big.\notag\\
&\qquad \quad \Big.+\{1+(e^{\lambda_i\tau_i}-1)^{\alpha_i}\}\{n_i+(n_i-k_i)(e^{\lambda_i\tau_i}-1)^{\alpha_i}\}\Big]
&&
\end{flalign}
\begin{flalign}
   \frac{\partial^2\big(ln\,L(\bm{\theta})\big)}{\partial b_{j_1}\partial b_{j_2}}&= \sum_{i=1}^{I}\frac{\alpha_i\lambda_i\tau_i x_{ij_1}x_{ij_2}e^{\lambda_i\tau_i}(e^{\lambda_i\tau_i}-1)^{-2}}{\{1+(e^{\lambda_i\tau_i}-1)^{\alpha_i}\}^2}\Big[k_i\alpha_i\lambda_i\tau_i e^{\lambda_i\tau_i}(e^{\lambda_i\tau_i}-1)^{\alpha}\Big.\notag\\
   &
   \quad\qquad+\{e^{\lambda_i\tau_i}-\tau_i\lambda_i-1\}\{1+(e^{\lambda_i\tau_i}-1)^{\alpha_i}\}\notag\\
   &\quad\qquad\qquad \Big.\{n_i+(n_i-k_i)(e^{\lambda_i\tau_i}-1)^{\alpha_i}\}\Big]
   &&
\end{flalign}
\begin{flalign}
\frac{\partial^2\big(ln\,L(\bm{\theta})\big)}{\partial a_{j_1}\partial b_{j_2}}&=\sum_{i=1}^{I}\frac{\alpha_i\lambda_i\tau_i x_{ij_1}x_{ij_2}e^{\lambda_i\tau_i}(e^{\lambda_i\tau_i}-1)^{-1}}{\{1+(e^{\lambda_i\tau_i}-1)^{\alpha_i}\}^2}\Big[k_i\alpha_i(e^{\lambda_i\tau_i}-1)^{\alpha_i}\Big.\notag\\
&\quad\qquad ln\,(e^{\lambda_i\tau_i}-1)+\{1+(e^{\lambda_i\tau_i}-1)^{\alpha_i}\}\notag\\
&\quad\qquad\qquad\Big.\{n_i+(n_i-k_i)(e^{\lambda_i\tau_i}-1)^{\alpha_i}\}\Big]
&&
\end{flalign}
where, $j_1,j_2=1,\dots,J$.
}
\section{Proof of Theorem \eqref{thm3}.}

Based on \citep{calvino2021robustness}, the proof of the theorem carries on as follows.\\

\noindent WDPD measure keeping only the terms involving parameters can be written as,
\begin{flalign*}
D_K&=\sum_{i=1}^{I}\frac{k_i}{K}\left[P^{\beta+1}_{i1}+\bar{P}^{\beta+1}_{i1}-\frac{\beta+1}{\beta}\left\{\frac{n_i}{k_i}P^{\beta}_{i1}+\left(\frac{k_i-n_i}{k_i}\right)\bar{P}^{\beta}_{i1}\right\}\right]\\
\text{where,}\; P_{i1}&=F(\tau_i;x_{ij},\bm{\theta})\;,\;\bar{P}_{i1}=1-P_{i1}=\bar{F}(\tau_i;x_{ij},\bm{\theta})\\
&&
\end{flalign*}
Let us define, $X_{ui}\sim Bin(1,P_{i1})$, then, $\sum_{ui=1}^{K_i}X_{ui}=n_i.$ Therefore,
\begin{flalign*}
 D_K&=\sum_{i=1}^{I}\frac{k_i}{K}\left[\frac{1}{k_i}\sum_{u_i=1}^{k_i}(P^{\beta+1}_{i1}+\bar{P}^{\beta+1}_{i1})\right.\\
&\qquad\qquad\qquad\left.-\frac{\beta+1}{\beta}\frac{1}{k_i}\left\{\sum_{u_i=1}^{k_i}X_{u_i}P^{\beta}_{i1}+\sum_{u_i=1}^{k_i}(1-X_{u_i})\bar{P}^{\beta}_{i1}\right\}\right]\\
&=\sum_{i=1}^{I}\frac{k_i}{K}\left[\frac{1}{k_i}\sum_{u_i=1}^{k_i}\left\{(P^{\beta+1}_{i1}+\bar{P}^{\beta+1}_{i1})\right.\right.\\
&\qquad\qquad\qquad\left.\left.-\frac{\beta+1}{\beta}\left(X_{u_i}P^{\beta}_{i1}+(1-X_{u_i})\bar{P}^{\beta}_{i1}\right)\right\}\right]\\
\implies D_K&=\sum_{i=1}^{I}\frac{k_i}{K}\left[\frac{1}{k_i}\sum_{u_i=1}^{k_i}H_{i\beta}(\bm{\theta})\right]
&&
\end{flalign*}
where,
\begin{flalign*}
H_{i\beta}(\bm{\theta})&=\left\{P^{\beta+1}_{i1}+\bar{P}^{\beta+1}_{i1}-\frac{\beta+1}{\beta}\left(X_{u_i}(P^{\beta}_{i1}-\bar{P}^{\beta}_{i1})+\bar{P}^{\beta}_{i1}\right)\right\}
&&
\end{flalign*}
For the $i^{th}$ group,
\begin{flalign*}
Y_i=\frac{\partial}{\partial\bm{\theta}}H_{i\beta}(\bm{\theta})&=(\beta+1)P^{\beta}_{i1}\frac{\partial(P_{i1})}{\partial\bm{\theta}}-(\beta+1)\bar{P}^{\beta}_{i1}\frac{\partial(P_{i1})}{\partial\bm{\theta}}-\frac{\beta+1}{\beta}\bigg[X_{ui}\bigg.\\
&\qquad\qquad\left.\left.\left(\beta P^{\beta-1}_{i1}\frac{\partial(P_{i1})}{\partial\bm{\theta}}\right.+\beta\bar{P}^{\beta-1}_{i1}\frac{\partial(P_{i1})}{\partial\bm{\theta}}\right)-\beta\bar{P}^{\beta-1}_{i1}\frac{\partial(P_{i1})}{\partial\bm{\theta}}\right]\\
&=(\beta+1)\left[P^{\beta}_{i1}-\bar{P}^{\beta}_{i1}-X_{ui}\left(P^{\beta-1}_{i1}+\bar{P}^{\beta-1}_{i1}\right)+\bar{P}^{\beta-1}_{i1}\right]\frac{\partial(P_{i1})}{\partial\bm{\theta}}\\
&=(\beta+1)\left[(P^{\beta-1}_{i1}+\bar{P}^{\beta-1}_{i1})(P_{i1}-X_{ui})\right]\frac{\partial}{\partial\bm{\theta}}P_{i1}.
\end{flalign*}
\begin{flalign*}
\text{Therefore,}\;\; Y_{is}&=(\beta+1)\left[(P^{\beta-1}_{i1}+\bar{P}^{\beta-1}_{i1})(P_{i1}-X_{ui})\right]\frac{\partial}{\partial\bm{\theta}_{s}}P_{i1}\\
\noindent \text{where}\qquad Y_i&=(Y_{i1},Y_{i2},\dots,Y_{is},\dots,Y_{2J})^{'}. 
&&
\end{flalign*}
\begin{flalign*}
\text{Here, }\quad E(Y_i)=0, \\
 Var(Y_{is})&=(\beta+1)^2\left[(P^{\beta-1}_{i1}+\bar{P}^{\beta-1}_{i1})\frac{\partial(P_{i1})}{\partial\bm{\theta}_s}\right]^2Var(X_{u_i})\\
&=(\beta+1)^2\left[(P^{\beta-1}_{i1}+\bar{P}^{\beta-1}_{i1})\frac{\partial(P_{i1})}{\partial\bm{\theta}_s}\right]^2 P_{i1}\bar{P}_{i1}
&&
\end{flalign*}
and,
\begin{flalign*}
Cov(Y_{is_1},Y_{is_2})&=\left[(\beta+1)(P^{\beta-1}_{i1}+\bar{P}^{\beta-1}_{i1})\right]^2\frac{\partial(P_{i1})}{\partial\bm{\theta}_{s_1}}\frac{\partial(P_{i1})}{\partial\bm{\theta}_{s_2}}Var(X_{u_i})\\
&=\left[(\beta+1)(P^{\beta-1}_{i1}+\bar{P}^{\beta-1}_{i1})\right]^2\frac{\partial(P_{i1})}{\partial\bm{\theta}_{s_1}}\frac{\partial(P_{i1})}{\partial\bm{\theta}_{s_2}}P_{i1}\bar{P}_{i1}\\
&&
\end{flalign*}
Therefore,
\begin{flalign*}
Cov\left(\frac{\partial}{\partial\bm{\theta}}H_{i\beta}(\bm{\theta})\right)&=(\beta+1)^2K_{i\beta}(\bm{\theta})\\
K_{i\beta}(\bm{\theta})_{ss}&=\frac{Var(Y_{is})}{(\beta+1)^2}\quad \text{(variance term)}\\
K_{i\beta}(\bm{\theta})_{s_1s_2}&=\frac{Cov(Y_{is_1},Y_{is_2})}{(\beta+1)^2}\quad \text{(Covariance term)}
&&
\end{flalign*}
Applying Central Limit Theorem (CLT), when $k_i \to \infty$ for $i=1,\dots,I,$
\begin{flalign*}
&\sqrt{k_i}\left(\frac{1}{k_i}\sum_{u_i=1}^{k_i}\frac{\partial}{\partial\bm{\theta}}H_{i\beta}(\bm{\theta})\right)\sim N\left(0,(\beta+1)^2K_{i\beta}(\bm{\theta})\right)\\
&&
\end{flalign*}
We denote,
\begin{flalign}
\nabla D_K(\bm{\theta})&=(D_{K,1}(\bm{\theta}),\dots,D_{K,s}(\bm{\theta}),\dots,D_{K,2J}(\bm{\theta}))^{'}\;\text{where,}\;\;D_{K,s}(\bm{\theta})=\frac{\partial}{\partial\bm{\theta}_s}D_K\notag\\
\text{define,}\;\;T_{\beta}&=-\sqrt{K}\nabla D_K(\bm{\theta})=-\sqrt{K}\sum_{i=1}^{I}\frac{k_i}{K}\left(\frac{1}{k_i}\sum_{u_i=1}^{k_i}\frac{\partial}{\partial\bm{\theta}}H_{i\beta}(\bm{\theta})\right)\notag\\
\implies T_{\beta}&\sim N\left(0,(\beta+1)^2\sum_{i=1}^{I}\frac{k_i}{K}K_{i\beta}(\bm{\theta})\right)\,,\,\\
&\text{where}\;\frac{k_i}{K}\;\text{is finite when}\;k_i\xrightarrow[]{}\infty\notag \label{a1}
&&
\end{flalign}
Now,
\begin{flalign*}
\frac{\partial(D_{K,s_1}(\bm{\theta}))}{\partial\bm{\theta}_{s_2}}&=\sum_{i=1}^{I}\frac{k_i}{K}\left(\frac{1}{k_i}\sum_{u_i=1}^{k_i}\frac{\partial^2H_{i\beta(\bm{\theta})}}{\partial\bm{\theta}_{s_2}\partial\bm{\theta}_{s_1}}\right)
&&
\end{flalign*}
where,
\begin{flalign*}
\frac{\partial^2H_{i\beta}(\bm{\theta})}{\partial\bm{\theta}_{s_2}\partial\bm{\theta}_{s_1}}
&=(\beta+1)\left[\left\{(P^{\beta-1}_{i1}+\bar{P}^{\beta-1}_{i1})(P_{i1}-X_{ui})\frac{\partial^2(P_{i1})}{\partial\bm{\theta}_{s_2}\bm{\theta}_{s_1}}\right\}\right.\\
&\quad\qquad\qquad+\left\{(P^{\beta-1}_{i1}+\bar{P}^{\beta-1}_{i1})+(P_{i1}-X_{ui})(\beta-1)\right.\\
&\qquad\qquad\qquad\left.\left.\left(P^{\beta-2}_{i1}-\bar{P}^{\beta-2}_{i1}\right)\right\}\frac{\partial(P_{i1})}{\partial\bm{\theta}_{s_2}}\frac{\partial(P_{i1})}{\partial\bm{\theta}_{s_1}}\right]\\
&&
\end{flalign*}
Applying CLT, when $k_i \to \infty$ for $i=1, \ldots,I,$ $\frac{1}{k_i}\sum_{u_i=1}^{k_i}X_{ui}\xrightarrow{p} P_{i1}$.
\begin{flalign*}
\noindent \text{Therefore,}\quad \frac{1}{k_i}\sum_{u_i=1}^{k_i}\frac{\partial^2H_{i\beta}(\bm{\theta})}{\partial\bm{\theta}_{s_2}\partial\bm{\theta}_{s_1}}\xrightarrow{p}(\beta+1)\left(P^{\beta-1}_{i1}+\bar{P}^{\beta-1}_{i1}\right)\frac{\partial(P_{i1})}{\partial\bm{\theta}_{s_2}}\frac{\partial(P_{i1})}{\partial\bm{\theta}_{s_1}}\\
\text{and,}\quad  \frac{\partial(D_{K,s_1}(\bm{\theta}))}{\partial\bm{\theta}_{s_2}}\xrightarrow{p}\sum_{i=1}^{I}\frac{k_i}{K}(\beta+1)\left(P^{\beta-1}_{i1}+\bar{P}^{\beta-1}_{i1}\right)\frac{\partial(P_{i1})}{\partial\bm{\theta}_{s_2}}\frac{\partial(P_{i1})}{\partial\bm{\theta}_{s_1}}
&&
\end{flalign*}
Consider $\bm{\theta}_{0}$ to be the true value of parameters, then apply the Taylor series expansion $D_{k,s}(\bm{\theta})$ around $\bm{\theta}_0$
\begin{flalign}
D_{K,s}(\bm{\theta})&=D_{K,s}(\bm{\theta}_0)+\sum_{m=1}^{2J}\left.\frac{\partial(D_{K,s}(\bm{\theta}))}{\partial\bm{\theta}_{m}}\right\vert_{\bm{\theta}=\bm{\theta}_0}(\hat{\bm{\theta}}_{m}-\bm{\theta}_{0m})\notag\\
&\quad+\frac{1}{2}\sum_{l=1}^{2J}\sum_{m=1}^{2J}\left.\frac{\partial^2 (D_{K,s}(\bm{\theta}))}{\partial\bm{\theta}_{l}\partial\bm{\theta}_{s}}\right\vert_{\bm{\theta}=\bm{\theta}^{\ast}}(\hat{\bm{\theta}}_{l}-\bm{\theta}_{0{l}})(\hat{\bm{\theta}}_{m}-\bm{\theta}_{0m})
\end{flalign}
Since, $D_{K,s}(\hat{\bm{\theta}}_{\beta})=0$  it can be written that,
\begin{flalign}
-\sqrt{K}D_{K,s}(\bm{\theta}_0)&=\sqrt{K}\sum_{m=1}^{2J}\left(\left.\frac{\partial(D_{K,s}(\bm{\theta}))}{\partial\bm{\theta}_{m}}\right\vert_{\bm{\theta}=\bm{\theta}_0}\right.\notag\\
&\qquad+\left.\frac{1}{2}\sum_{l=1}^{2J}\left.\frac{\partial^2(D_{K,s}(\bm{\theta}))}{\partial\bm{\theta}_{l}\partial\bm{\theta}_{m}}\right\vert_{\bm{\theta}=\bm{\theta}^{*}}(\hat{\bm{\theta}}_{l}-\bm{\theta}_{0{l}})\right)  (\hat{\bm{\theta}}_{m}-\bm{\theta}_{0m}) \label{a2}
\end{flalign}
Denoting,
\begin{flalign*}
A_{s,m}&=\left.\frac{\partial(D_{K,s}(\bm{\theta}))}{\partial\bm{\theta}_{m}}\right\vert_{\bm{\theta}=\bm{\theta}_0}+\frac{1}{2}\sum_{l=1}^{2J}\left.\frac{\partial^2(D_{K,s}(\bm{\theta}))}{\partial\bm{\theta}_{l}\partial\bm{\theta}_{m}}\right\vert_{\bm{\theta}=\bm{\theta}^{*}}(\hat{\bm{\theta}}_{l}-\bm{\theta}_{0{l}})
&&
\end{flalign*}
Then,
\begin{flalign}
A_{s,m}&\xrightarrow{p}\sum_{i=1}^{I}\frac{k_i}{K}(\beta+1)\left(P^{\beta-1}_{i1}+\bar{P}^{\beta-1}_{i1}\right)\frac{\partial(P_{i1})}{\partial\bm{\theta}_{m}}\frac{\partial(P_{i1})}{\partial\bm{\theta}_{s}}\notag\\
\text{where}&\;\frac{k_i}{K}\;\text{is finite when}\;k_i\xrightarrow[]{}\infty. \notag\\
\implies& A_{\beta}\xrightarrow{p}(\beta+1)J_{\beta}(\bm{\theta}_0)
\label{a3}
&&
\end{flalign}
where,$\;A_{\beta}$ is the $2J\times 2J$ matrix with $(s,m)$th element as $A_{s,m}$ and
\begin{flalign*}
J_{\beta}(\bm{\theta}_0)&=\left[\left(\sum_{i=1}^{I}\frac{k_i}{K}\left(P^{\beta-1}_{i1}+\bar{P}^{\beta-1}_{i1}\right)\frac{\partial(P_{i1})}{\partial\bm{\theta}_{m}}\frac{\partial(P_{i1})}{\partial\bm{\theta}_{s}}\right)_{s,m}\right]
&&
\end{flalign*}
We can write, 
\begin{flalign*}
-\sqrt{K}D_{K,s}(\bm{\theta}_0)&=\sqrt{K}  \sum_{s=1}^{2J} (\hat{\bm{\theta}}_{s}-\bm{\theta}_{0s})A_{s,m}
\end{flalign*}
Therefore, it can be expressed as
\begin{flalign*}
&T_{\beta}=\sqrt{K}(\hat{\bm{\theta}}_{\beta}-\bm{\theta}_{0})A_{\beta}\\
&\implies \sqrt{K}(\hat{\bm{\theta}}_{\beta}-\bm{\theta}_{0}) =A_{\beta}^{-1}T_{\beta}\\
&\implies\sqrt{K}(\hat{\bm{\theta}}_{\beta}-\bm{\theta}_0)=A_{\beta}^{-1}T_{\beta}\sim N\left(0,J_{\beta}(\bm{\theta}_0)^{-1}K_{\beta}(\bm{\theta}_0)J_{\beta}(\bm{\theta}_0)^{-1}\right)\\
&\qquad\qquad\qquad\qquad\qquad\qquad\qquad\qquad\qquad\qquad(\text{by equation \eqref{a1}})
&&
\end{flalign*}
where,
\begin{flalign*}
K_{\beta}(\bm{\theta}_0)&=\sum_{i=1}^{I}\frac{k_i}{K}K_{i\beta}(\bm{\theta})\;;\;\frac{k_i}{K}\;\text{is finite, when}\;k_i\xrightarrow[]{}\infty.
&&
\end{flalign*}

\bibliographystyle{elsarticle-num} 
 \bibliography{cas-refs}

\end{document}